\documentclass[twocolumn]{aastex63}

\usepackage{amsmath}
\usepackage{bm}

\bmdefine{\by}{y}
\bmdefine{\bs}{s}
\bmdefine{\mm}{m}
\bmdefine{\btheta}{\theta}
\bmdefine{\bphi}{\phi}
\bmdefine{\bt}{t}
\bmdefine{\bB}{B}

\received{May 21, 2021}
\revised{September 8, 2021}
\accepted{September 24, 2021}
\submitjournal{AJ}

\shorttitle{ParSNIP: Generative Models of Transient Light Curves}
\shortauthors{Boone}
\graphicspath{{./}{figures/}{figures_architecture/}}
\begin{document}

\title{ParSNIP: Generative Models of Transient Light Curves with Physics-Enabled Deep Learning}

\correspondingauthor{Kyle Boone}
\email{kyboone@uw.edu}

\author[0000-0002-5828-6211]{Kyle Boone}
\affiliation{DiRAC Institute, Department of Astronomy, University of Washington, 3910 15th Ave NE, Seattle, WA, 98195, USA}

% \author{et al.}
% \nocollaboration{1}



\begin{abstract}

We present a novel method to produce empirical generative models
of all kinds of astronomical transients from datasets of
unlabeled light curves. Our hybrid model, that we call ParSNIP, uses a neural network
to model the unknown intrinsic diversity of different transients and
an explicit physics-based model of how light from the transient propagates
through the universe and is observed. 
The ParSNIP model predicts the time-varying spectra of transients
despite only being trained on photometric observations. With a three-dimensional
intrinsic model, we are able to fit out-of-sample multiband light curves of many
different kinds of transients with model uncertainties of 0.04--0.06~mag.
The representation learned by the ParSNIP model is invariant to redshift,
so it can be used to perform photometric classification of transients
even with heavily biased training sets. Our classification techniques significantly
outperform state-of-the-art methods on both simulated (PLAsTiCC)
and real (PS1) datasets with 2.3$\times$ and 2$\times$ less contamination respectively
for classification of Type~Ia supernovae.
We demonstrate how our model can identify previously-unobserved kinds
of transients and produce a sample that is 90\% pure. The ParSNIP model can
also estimate distances to Type~Ia supernovae in the PS1 dataset with an
RMS of $0.150 \pm 0.007$~mag compared to $0.155 \pm 0.008$~mag for the
SALT2 model on the same sample. We discuss how our model could
be used to produce distance estimates for supernova cosmology without
the need for explicit classification.

\end{abstract}



\section{Introduction} \label{sec:introduction}

Upcoming surveys such as the Legacy Survey of Space and Time
\citep[LSST;][]{ivezic19}
that will be conducted at the Vera C. Rubin Observatory, and
the Nancy Grace Roman Space Telescope \citep{spergel15}
will obtain light curves for millions
of astronomical transients. These large datasets will enable many
novel science applications. The number of observed Type~Ia supernovae
(SNe~Ia) will increase by more than two orders of magnitude compared
to current samples \citep{lsst09, scolnic18}. These large samples of SNe~Ia
will be used to constrain cosmological parameters both by improving measurements
of the expansion history of the universe and by performing novel
measurements of local peculiar velocity field \citep{graziani20}.
These surveys will also observe large numbers of
other kinds of transients, such as superluminous supernovae (SLSN), which
will be helpful both to elucidate the origins of these transients
\citep{villar18} and to use them as new cosmological probes \citep{scovacricchi16}.
It is also likely that new kinds of rare transients that
have not previously been observed will be discovered in the datasets
from these surveys.

All of these science applications rely on being able to extract information
from light curves and distinguish between light curves of different kinds
of transients. Observations can be compared directly to simulations
and theory \citep{kerzendorf21}, but these models are not currently accurate enough
to characterize observed light curves at the precision required for many science applications.
For SNe~Ia, a range of different empirical models have been created
to describe their light curves and spectra \citep{guy07, saunders18, leget20, mandel20}.
These are generative models in that they predict the full time-varying
spectrum of a transient in terms of a small number of parameters
whose distribution is reasonably well known.

These models all include explicit descriptions of how the light propagates through the universe
and is observed on a detector. Propagation effects include redshifting of the light from the
transient, dust along the line of sight, and luminosity variation due to effects such as 
weak gravitational lensing. A light curve can be observed with different
cadences, with different telescopes, in different bandpasses, at different
signal-to-noises, or at different times. All of these effects will change the observed
light curve for a transient, but they do not affect the intrinsic physics of the underlying transient.
We therefore refer to them as ``symmetries''. By explicitly
handling these symmetries, the previously-described models of SNe~Ia
obtain consistent values of the model parameters when fitting light curves observed
under different conditions. In particular, these models can obtain consistent estimates
of the luminosity when fitting light curves at different redshifts which is essential
for estimating distances to SNe~Ia.

Similar models do not exist for other kinds of transients. Instead,
previous work using light curves of non-SNe~Ias has typically focused on
extracting features from the light curves rather than building a generative
model. Features can be extracted by fitting an empirical functional
form to the light curve \citep{bazin09, sanders15}, by fitting parameters
of a theoretical model to match the light curve \citep{guillochon18},
by estimating a smooth approximation to the light curve from which
arbitrary empirical features can be measured \citep{lochner16, boone19, alves21},
or by using a neural network \citep{muthukrishna19, moller19}.
These methods produce a
large number of features that can be used to characterize a given
light curve, but these features are not typically invariant to
the symmetries of how the light curve was observed. In particular, the features
produced by all of these models tend to be highly dependent on redshift.
This is a major challenge for science applications such as photometric
classification where labeled datasets tend to be heavily biased towards
bright, low-redshift transients \citep{lochner16, boone19}.

In this work, we address these challenges by building a generative model
of all transient light curves that is invariant to observing conditions.
Most techniques that have previously been used to build generative models of SNe~Ia assume
that the diversity of SNe~Ia can be described by a simple linear model.
Such models provide a good description of most normal SNe~Ia, but they
fail to describe peculiar SNe~Ia \citep{boone21a, boone21b} and are
not flexible enough to describe most other kinds of transients.
Furthermore, the techniques used to build these models require large
datasets of well-measured light curves and spectra \citep{betoule14}
that are not available for transients other than SNe~Ia.

Instead, we learn a generative model directly from large samples of
light curves using a modified version of a variational autoencoder
\citep[VAE;][]{kingma14}. A VAE learns a low-dimensional representation
of its input that we call a latent space.
First, an encoder model uses variational inference to approximate the
posterior distribution over the latent space for a given input.
A generative model, also referred to as a decoder, then reconstructs the input given a
point in the latent space. The VAE can be trained by applying
both the encoder and decoder to a given input and by comparing the
original input to the reconstructed one.

Autoencoders have previously been applied to astronomical light curves
\citep{naul18, pasquet19, martinezpalomera20, villar20, villar21} and shown to be effective
for tasks such as photometric classification and outlier identification.
However, these models do not include explicit descriptions of observing symmetries,
so the same transient observed under different conditions (e.g. redshift) will
be assigned to different locations in the latent space. In this work,
we produce a hybrid physics-VAE model where we use a neural network to describe
the intrinsic time-varying spectra of astronomical transients and an explicit physical model for
known symmetries. Our model produces a three-dimensional representation of
the intrinsic diversity
of transients that is insensitive to observing conditions, including redshift.
We call the resulting model ``ParSNIP'' (Parameterization of SuperNova Intrinsic Properties).

We describe the datasets that we use for this work in Section~\ref{sec:dataset}.
In Section~\ref{sec:methods}, we describe the ParSNIP model and how it is trained.
We evaluate the performance of the ParSNIP model as a generative model in
Section~\ref{sec:performance} and show how it is able to learn the full
time-varying spectra for transients despite only being trained on light curve data.
In Section~\ref{sec:applications}, we show how the ParSNIP model achieves
state-of-the-art performance for tasks
such as photometric classification, identification of new transients, or cosmological
distance estimation. Finally,
in Section~\ref{sec:discussion} we discuss how the ParSNIP model can be applied
to other datasets and how it could be improved.

\section{Dataset} \label{sec:dataset}

We evaluate the techniques described in this work on two different datasets of
supernova-like light curves. First, we use a dataset of observed supernova-like
light curves from the Pan-STARRS1 Medium-Deep Survey (PS1) \citep{chambers16},
the details of which
are described in \citet{villar20}. This dataset contains 2,885 light curves
with host-galaxy redshifts of which 557 have spectroscopically-confirmed types.
The light curves were observed in the Pan-STARRS g, r, i, and z bandpasses with
a cadence of $\sim$7 days per filter.

We also evaluate the ParSNIP model on the PLAsTiCC dataset of simulated light curves
described in \citep{kessler19}. This dataset contains 3,006,109 simulated supernova-like light
curves with observations in the Rubin Observatory's u, g, r, i, z, and y bandpasses,
and the authors aimed to simulate a realistic dataset of light curves from
three years of operation of the Rubin Observatory.
This simulation consists of two distinct surveys. The vast majority of the light
curves (2,972,316) are from the Wide-Fast-Deep (WFD) survey which covers around
half of the sky
with observations roughly twice per week in at least one of the six bandpasses.
33,793 light curves are from the Deep-Drilling-Fields (DDF) survey which covers
$\sim$50 deg$^2$ with significantly deeper and more frequent observations.
The authors of the PLAsTiCC dataset also simulated a spectroscopic followup
strategy which produces type labels for 5,153 of the light curves in this dataset.
This labeled dataset is highly biased towards bright transients, and has a mean
redshift of 0.32 compared to 0.50 for the full dataset.

For this analysis, we assume that the redshifts of all of the transients
are known. This assumption is not required for our analysis, but it does simplify
our methodology. We will discuss how our methods can be applied to datasets
where the redshift is not available in Section~\ref{sec:photoz}.

We restrict our analysis to supernova-like light curves. By this, we mean
that we consider any extragalactic light curves where there is an excess of flux for a
single well-defined time period, and where the light curve has a stable
background level before and after this time period. The PS1 dataset that
we use was already restricted to these kinds of light curves. For the PLAsTiCC
dataset, we reject all of the galactic and AGN light curves but keep all other
kinds of transients.

\subsection{Preprocessing} \label{sec:preprocessing}

To preprocess our light curves, we first make a rough estimate of the time of maximum light of the light curve
by taking the median time of the five highest signal-to-noise observations in the
light curve. In each bandpass, we estimate the background level using a biweight estimator
\citep{beers90} on all observations at least 250 days from our estimated time of
maximum light. We subtract this estimate of the background level from each
light curve. For the PS1 dataset, we also correct the light curves for Milky Way
extinction using the dust map from \citet{schlafly11}.

Neural networks typically perform better if their inputs have a limited range,
but flux values can vary over many orders of magnitude. To address this,
we normalize the brightness of each light curve by its maximum flux in
observations in any band with a signal-to-noise of at least five.
After this procedure, most
flux values lie between zero and one. We record the normalization scale, and
use it in future analyses such as distance estimation.

We find that the statistical uncertainties that are reported for both the PLAsTiCC
and PS1 datasets heavily underestimate the true uncertainties for very high
signal-to-noise observations. This is partially due to uncertainties in
our background subtraction procedure. To prevent these underestimated
uncertainties from having excessively large weights in our model, we add
an error floor of 0.01 to all of our flux uncertainties when training the model.
With our normalization, this roughly corresponds to an error floor of 1\% of the peak flux.

Astronomical light curves are typically sparsely sampled, but most
neural network architectures require inputs that are sampled on an
evenly-spaced grid. To be able to use these architectures, we evaluate
our light curves on a grid following a procedure similar to the
one described in \citet{pasquet19}. Note that astronomical observations
from a given telescope typically only occur during a short time interval
each night when the target is near the zenith. All observations of the
same target will be separated in time by some integer number
of sidereal days with residuals of only $\sim$1 hour. We can therefore
evaluate the light curve on a grid of sidereal days and preserve almost all of
the temporal information.

For each transient, we build a grid of observations that has a length
of 300 sidereal days centered on the estimated time of maximum light.
The grid has rows for the flux and inverse flux variance in each bandpass.
For nights where the transient was observed in a given bandpass, we include the observed
flux value and inverse flux variance in the grid. For nights when the transient was
not observed, we simply input zero for both the flux and inverse flux variance. We also include
one row in the grid for the redshift, with the same value repeated at all times. After this
procedure, our light curves are represented by a two-dimensional grid of size
$300\times (2 N + 1)$ where $N$ is the number of bandpasses that we are modeling.
We can then use standard neural network architectures to process our light
curves.

\section{Methods} \label{sec:methods}

\subsection{Overview}

The overall design goal for our model is to build a representation of the
diversity of transients that disentangles intrinsic diversity related
to the physics of the transient from diversity related to the symmetries
of how the transient was observed. A transient should be assigned
the same intrinsic representation regardless of how it was observed.
In particular, we would like for out model to be invariant to redshift, meaning
that a transient is assigned the same intrinsic representation
regardless of the redshift that it is observed at.

Many different approaches exist for modeling disentangled representations with
autoencoders \citep{tschannen18}. These approaches mostly rely on either
using a supervised training procedure (which would require labels that we do not have)
\citep[e.g.][]{kulkarni15} or using regularization to encourage an implicit disentangled
representation \citep[e.g.][]{higgins17}. We adopt an approach similar
to that of spatial-VAE \citep{bepler19} where we generate a disentangled representation
by explicitly including known symmetries in our generative model.

\begin{figure*}
\plotone{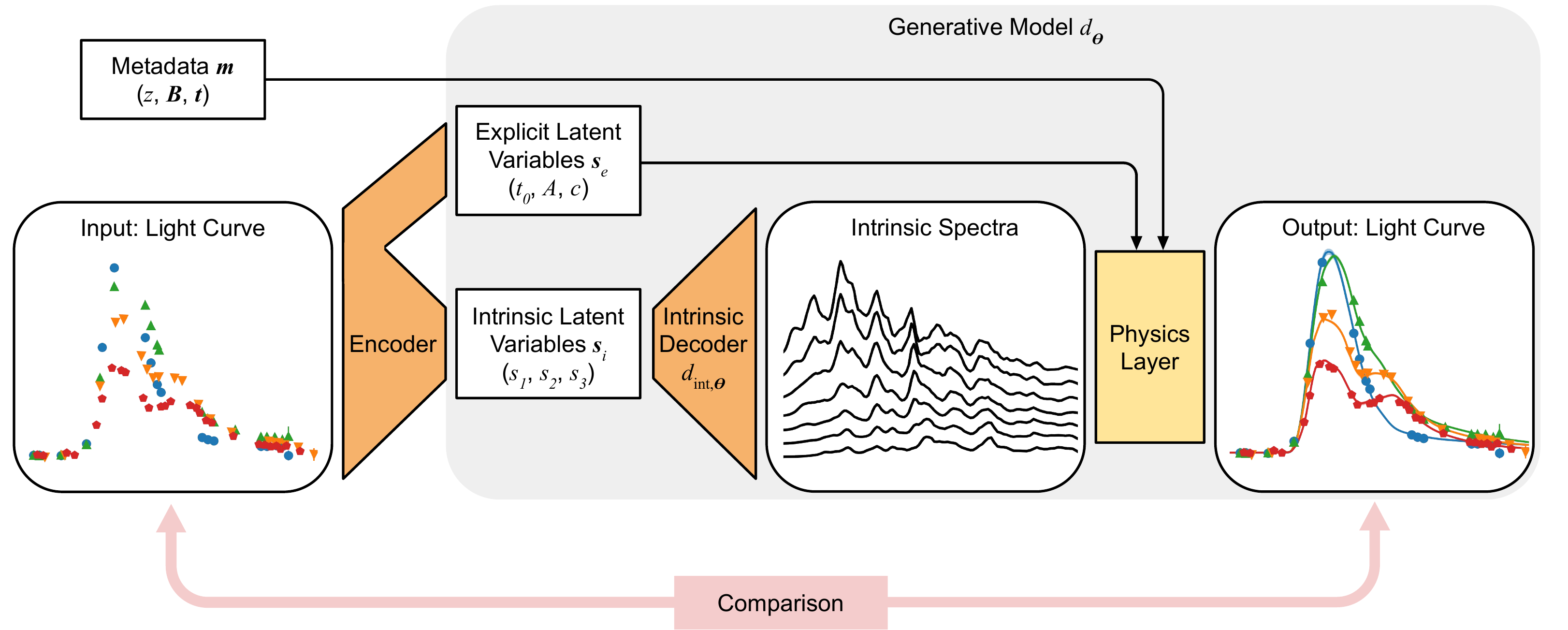}
\caption{
    Schematic overview of the ParSNIP model. An encoder model predicts the posterior distribution over
    a set of latent variables for a given input light curve. We label a subset of these
    latent variables as intrinsic latent variables, and use them as input to an intrinsic decoder
    model that predicts the full intrinsic time-varying spectrum of the transient. A physics layer
    then describes how light propagates through the universe and is observed on a detector.
    The physics layer takes as input a set of explicit latent variables and metadata
    describing the observing conditions. The encoder and intrinsic decoder are implemented as
    neural networks. The physics layer uses only physical relations with known
    functional forms and it has no free model parameters.
}
\label{fig:schematic}
\end{figure*}

Our generative model first uses a neural network to predict the intrinsic spectra of a given transient
as a function of three intrinsic latent variables. The intrinsic spectra are then passed
through a physics layer that explicitly models how the light propagates through the universe
and is observed. The physics layer also takes as input a set of explicit latent
variables and metadata describing the observing conditions. A schematic overview of our model
is shown in Figure~\ref{fig:schematic}.

\subsection{Variational Autoencoders} \label{sec:vae}

To build this model, we developed a modified version of a VAE \citep{kingma14}.
As traditionally formulated, a VAE
is used to model some observed random variables $\by$ in terms of some
unobserved random variables $\bs$. We assume that $\bs$ are drawn from some
prior distribution
$p_\btheta(\bs)$, and that $\by$ are then drawn from
the conditional distribution $p_\btheta(\by|\bs)$.
These distributions are assumed to be part of some family of
distributions parameterized by the hyperparameters $\btheta$.

The posterior distribution $p_\btheta(\bs|\by)$
is in general intractable. Instead of modeling it directly, in a
VAE we instead assume that the posterior can be
parameterized using some family of distributions $q_\bphi(\bs|\by)$
with hyperparameters $\bphi$. As shown in \citet{kingma14}, the
marginal likelihood of a given observation $\by$ is then bounded by:
\begin{align}
    % \log p_\btheta(\by) &= D_{KL}(q_\bphi(\bs|\by) || p_\btheta(\bs|\by)) + \mathcal{L}(\btheta, \bphi; \by) \nonumber \\
    % & \geq \mathcal{L}(\btheta, \bphi; \by) \nonumber \\
    % & \geq \mathbb{E}_{q_\bphi(\bs|\by)}\left[\log p_\btheta(\by,\bs) -\log q_\bphi(\bs|\by) \right] \nonumber \\
    % & \geq  \mathbb{E}_{q_\bphi(\bs|\by)}\left[ \log p_\btheta(\by|\bs) \right] \nonumber \\
    % & \indent - D_{KL}(q_\bphi (\bs|\by) || p_\btheta(\bs))
    \log p_\btheta(\by) & \geq \mathbb{E}_{q_\bphi(\bs|\by)}\left[ \log p_\btheta(\by|\bs) \right] \nonumber \\
    & \indent - D_{KL}(q_\bphi (\bs|\by) || p_\btheta(\bs))
    \label{eq:elbo}
\end{align}

The right side of this equation is referred to as the ``evidence lower bound'' or ELBO.
The first term of the ELBO captures the likelihood of the data under our model, and can
be interpreted as the reconstruction error of the model. The second term represents the
Kullback-Leibler divergence \citep{kullback51} between the approximate posterior
$q_\bphi(\bs|\by)$ and the prior $p_\btheta(\bs)$, and
is effectively a constraint on the form of the approximate posterior.

Typically, an isotropic multivariate Gaussian is assumed for the prior over the latent
representation $\bs$:
\begin{align}
    p_\btheta(\bs) = \mathcal{N}(\bm{0}, \bm{I})
\end{align}
and $p_\btheta(\by|\bs)$ is assumed to be described by a Gaussian distribution
with a diagonal covariance matrix whose mean is given by some deterministic function
$d_\theta(\bs)$:
\begin{align}
    p_\btheta(\by|\bs) = \mathcal{N}(d_\btheta(\bs), \sigma_\by^2 \bm{I})
\end{align}

Finally, we assume that $q_\bphi(\bs|\by)$ is described by a Gaussian distribution
with a diagonal covariance matrix:
\begin{align}
    \label{eq:encoder}
    q_\bphi(\bs|\by) = \mathcal{N}(\mu_\bphi(\by), \sigma_\bphi(\by)^2 \bm{I})
\end{align}
where $\mu_\bphi(\by)$ and $\sigma_\bphi(\by)$ are some deterministic functions
of $\by$. The validity of this assumption will be discussed in
Section~\ref{sec:posteriors}.

Training a VAE then consists of finding appropriate functions for $d_\btheta(\bs)$,
$\mu_\bphi(\by)$, and $\sigma_\bphi(\by)$ that maximize the ELBO in
Equation~\ref{eq:elbo}. These functions are typically implemented as neural
networks. A VAE has an architecture that resembles a classical autoencoder.
The function $\mu_\bphi(\by)$ is analogous to the encoder that takes a set of
observations $\by$ and outputs a location in the latent representation $\bs$. $d_\btheta(\bs)$
can be interpreted as a decoder that generates a forward model of the observations
from the latent representation. The full decoder model of the VAE given by
\begin{align}
    p_\btheta(\by,\bs) = p_\btheta(\by|\bs) p_\btheta(\bs)
\end{align}
is a generative model.

\subsection{Incorporating Symmetries into Variational Autoencoders} \label{sec:vae_symmetries}

In this work we modify the standard VAE architecture to
explicitly incorporate symmetries into our model. Note that the functions $d_\btheta(\bs)$,
$\mu_\bphi(\by)$, and $\sigma_\bphi(\by)$ are arbitrary functions. Rather than implementing
these purely as neural networks, we instead use a hybrid implementation where we explicitly
model symmetries of the observations where the functional form is known, and where
we use neural networks for parts of the model whose functional forms are a priori unknown.

Formally, we identify the subset of latent variables that represent known symmetries
of the model as ``explicit latent variables'', and we label them $\bs_e$.
We call the remaining latent variables that represent diversity whose functional
form is unknown ``intrinsic latent variables'', and we label them $\bs_i$.
Then $\bs = \{\bs_e, \bs_i\}$. We also allow for additional
metadata of the observations $\mm$ that describe known observing conditions
(e.g. the time that an observation was taken at). With these definitions,
we model the decoder as
\begin{align}
    d_\btheta(\bs_i, \bs_e, \mm) = f_1(\bs_e, \mm, d_{\text{int},\btheta}(f_2(\bs_e, \bs_i, \mm)))
\end{align}
where $d_{\text{int},\btheta}$ is an arbitrary function that models the intrinsic diversity.
$f_1$ and $f_2$ are fixed functions that apply known
symmetries to the inputs and outputs of $d_{\text{int},\btheta}$.

To model transient light curves, we choose to use explicit functional forms for the amplitude
of the light curve $A$, the color (capturing dust reddening of the light curve) $c$,
and a reference time for the light curve $t_0$. Each observation
also has associated metadata for the redshift of the light curve $z$,
the time of the observation $\bt$, and the bandpass that was used for the observation
$\bB(\lambda)$. In our formalism, we have $\bs_e = \{A, c, t_0\}$ and
$\mm = \{z, \bt, \bB(\lambda)\}$. We incorporate the known effects of each of these terms
on astronomical observations into the $f_1$ and $f_2$ functions to produce the following
decoder
\begin{align}
    \label{eq:decoder}
    & d_\btheta(\bs_i, A, c, t_0, z, \bt, \bB(\lambda)) \\
    \nonumber & = \int d\lambda \cdot A \cdot \bB(\lambda) \cdot C(c, \frac{\lambda}{1 + z}) \cdot d_{\text{int},\btheta}\left(\frac{\bt - t_0}{1 + z}, \frac{\lambda}{1 + z}, \bs_i\right)
\end{align}
where $C(c, \lambda)$ is a fixed dust extinction law that is applied in the rest frame, and
$\bB(\lambda)$ are the bandpasses used for each observation with known throughput.
We choose to use the color law $C(c, \lambda)$ from \citet{fitzpatrick07} in this
work with $R_V = 3.1$ as implemented in the \texttt{extinction} package \citep{barbary16b}.
$d_{\text{int},\btheta}$ is then effectively modeling the time-evolution of the rest frame
spectrum of the transient.

The SALT2 model that is often used to fit SNe~Ia \citep{guy07} is
actually a special case of the general model described in Equation~\ref{eq:decoder}
with the a single intrinsic latent parameter $x_1$ describing the width of the light curve,
i.e. $\bs_i = \{x_1\}$. For SALT2, the intrinsic diversity is modeled by a linear sum
of two spline template functions $M_0(\lambda, t)$ and $M_1(\lambda, t)$:
\begin{align}
    d_{\text{int},\text{SALT2}}(\bt, \lambda, x_1) = M_0(\lambda, \bt) + x_1 M_1(\lambda, \bt)
\end{align}
The SALT2 model provides a good description of most SNe~Ia, but its model of intrinsic
diversity is linear which is insufficient for some SNe~Ia \citep{boone21a, boone21b}
and most other kinds of transients.

\subsection{Modeling Spectra with a Neural Network}

In order to model a wide range of transients, we choose to implement
$d_{\text{int},\btheta}$ as a multilayer perceptron (MLP). This
function takes as input both the time of an observation and the intrinsic latent variables
$\bs_i$. A large part of the spectrum is typically needed to compute bandpass
photometry, so for computational reasons we choose to predict the full restframe spectrum with
the MLP on a grid of wavelength elements rather than having the wavelength as an additional input. 
Synthetic photometry can then be computed by numerically evaluating the integral in
Equation~\ref{eq:decoder}. Labeling the index of each restframe wavelength bin as $n$, our
implementation of the decoder can then be written as
\begin{align}
    & d_\btheta(\bs_i, A, c, t_0, z, \bt, \bB(\lambda)) \\
    \nonumber & = \sum_{n} A \cdot \bB'_n \cdot C(c, \lambda'_n) \cdot d_{\text{int},\btheta,n}\left(\frac{\bt - t_0}{1 + z}, \bs_i\right)
\end{align}
Here $\lambda'_n$ is the central rest-frame wavelength for bin $n$. $\bB'_n$ are the
integrated bandpasses for bin $n$ which can be computed by integrating the bandpasses
over each wavelength bin
\begin{align}
    \label{eq:integrated_bandpasses}
    \bB'_n = \int_{\lambda'_{n,\text{min}}}^{\lambda'_{n,\text{max}}} d\lambda \cdot \bB((1 + z) \lambda)
\end{align}

We choose to use a grid of 300 spectral elements with rest frame wavelengths
that are logarithmically-spaced between 1,000 and 11,000~\AA. 
Logarithmically-spaced wavelength elements simplify computing the integrated bandpasses
$\bB'_n$ because redshifting the bandpass is equivalent to a translation of the wavelength
elements. We precompute the integrated bandpasses $\bB'_n$ by numerically evaluating the
integral in Equation~\ref{eq:integrated_bandpasses} with 51 times oversampling.
With this configuration, we find that the numerical uncertainties in our synthetic photometry
are $\sim$0.0002~mag which is much smaller than typical measurement uncertainties.

To summarize, we implement $d_{\text{int},\btheta}$ as an MLP that takes as inputs
the intrinsic latent variables $\bs_i$ and a time.
The output of this MLP is a restframe spectrum on a grid of 300 spectral elements.
We then use an explicit physics model of host-galaxy dust,
redshift, and the bandpasses of the telescope to calculate synthetic photometry from
the predicted spectrum and compare with observations. The MLP has several hyperparameters,
namely the number of layers, width of each layer, and
activation function. The values of these hyperparameters will be discussed in
Section~\ref{sec:hyperparameters}.

With this architecture, the full decoder model $d_{\btheta}$ naturally handles all
of the previously-described symmetries of how the transient was observed. We can evaluate
the full model in different bandpasses or at different redshifts without making
any changes to the intrinsic model or to the intrinsic latent variables $\bs_i$.
This is a major difference from previous autoencoder models that predict the
light curve in specific bands directly, and that therefore have a representation
where the intrinsic properties are entangled with properties of how the observations
were taken \citep[e.g.][]{pasquet19, villar20}.

\subsection{Symmetry-Aware Encoders}

We choose to model the posterior distribution of the latent variables for a given light
curve as a Gaussian distribution with a diagonal covariance matrix, as shown in Equation~\ref{eq:encoder}. We assume a
$\mathcal{N}(\bm{0}, \bm{I})$ prior for the intrinsic latent variables. For the
color $c$, we assume a weak prior of $\mathcal{N}(0, 0.3^2)$ which is much larger than
the true dust distribution. Similarly, we assume a weak prior of $\mathcal{N}(0, (20~\text{days})^2)$
for the reference time $t_0$ which should be much wider than the uncertainty from the
estimated times of maximum light from the preprocessing step in Section~\ref{sec:preprocessing}.
As will be discussed in Section~\ref{sec:marginalizingamplitude}, we marginalize over the
amplitude and evaluate its posterior directly so it is ultimately not predicted by our encoder.

\begin{figure}
\epsscale{0.5}
\plotone{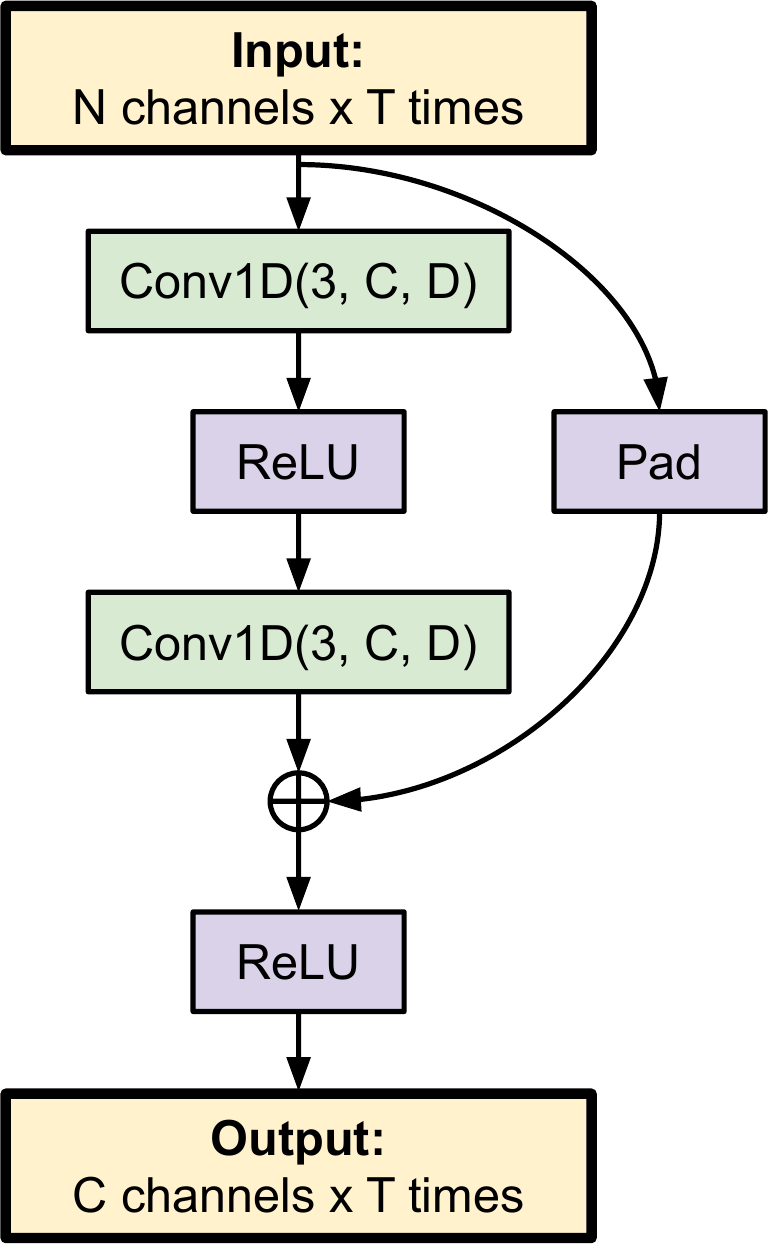}
\caption{
    Architecture of the residual blocks used for the encoder of our model. Each
    block takes as input a 2D array consisting of $N$ channels at $T$ times. Two one-dimensional
    convolutional layers are applied along the time dimension with a kernel size of 3 and a dilation of $D$
    elements. The input is then padded with zeros in the channel dimension and added with the output of the
    second convolutional layer to produce a 2D output with $C$ channels at $T$ times. We label this
    block as ``ResidualBlock(D, C)''.
}
\label{fig:residual_block}
\end{figure}

We implement the encoder using residual blocks with convolutional layers
\citep{he16}. Each block has two convolutional layers with a kernel size of 3, and we
use ReLU activation functions. A schematic of each block is shown in
Figure~\ref{fig:residual_block}. We chain a series of seven these blocks together with
increasing dilation sizes so that after all seven blocks the receptive field of the network
contains the full light curve. In a traditional VAE, the posteriors of the
latent variables would be predicted
using fully connected layers. While such a network should be theoretically capable of
representing the encoding functions for our model, in
practice, we find that these networks are very unstable and difficult to train. To address
this challenge, we add custom layers that are aware of the symmetries in our data
to predict the explicit latent variables.

\subsection{A Time-Invariant Encoder}

The main challenge for convergence is computing the reference time $t_0$ for a light
curve. At the start of training, the decoder will not be producing functions that look light
light curves, so the gradient of the loss function with respect to $t_0$ will not be meaningful
and the training will progress very slowly if at all.
We address this by constructing a novel neural network architecture that is mathematically
invariant to time translation.

The encoder predicts the hyperparameters $\mu$ and $\sigma^2$ describing the
posterior distributions over the latent variables in our model. To predict the mean reference time $\mu_{t_0}$, we
require our encoder to be a time-equivariant function, i.e. a function $f(t)$ that satisfies
\begin{align}
    f(t + \Delta t) = f(t) + \Delta t
\end{align}
For all of the other model hyperparameters, we require our encoder to be a time-invariant
function, i.e. one that satisfies:
\begin{align}
    f(t + \Delta t) = f(t)
\end{align}

Convolutional layers on their own are equivariant:
if the input to a convolutional layer is shifted, the output is identical but shifted by the same
amount as the input. By applying a series of the residual blocks shown in
Figure~\ref{fig:residual_block} without applying any pooling operations, we obtain an output
that is mathematically equivariant to any shift in the input light curve by an integer number of days.
The output of this layer is a two-dimensional array that has a length that is identical to the
length of the input light curve (in our case 300 days) and 200 channels.

To produce a time-invariant function, we throw away all of the time information by applying a
global max-pooling layer to this array.
This operation involves taking the maximum value of the array over the time dimension to collapse it in the
time dimension and produce a one-dimensional array with 200 channels. We use
fully-connected layers over the collapsed array to predict all of the hyperparameters $\mu$ and $\sigma^2$
describing the posterior distributions over the latent variables in our model except for the mean
reference time $\mu_{t_0}$. This procedure is mathematically invariant to any shift in time
of the input light curves by an integer number of days.

To product a time-equivariant function, we begin with the previously-discussed (300 days) $\times$ (200 channels) array
that was output from the residual blocks. We use a fully-connected layer applied to each bin in the
time dimension individually to collapse the channel dimension and obtain a one-dimensional vector $v$ with a
single value for each of the 300 time bins. We then apply a softmax
transformation to this vector to obtain a 300-dimensional vector that sums to one. The entries in this
vector can be thought of as weights for whether $\mu_{t_0}$ appears in a specific bin. We predict the
value of $\mu_{t_0}$ by taking the dot product of this vector with the index of each element in the vector
(i.e. a vector of increasing integers from 1 to 300).
\begin{align}
    \mu_{t_0} = \text{SoftMax}[v] \times \text{Index}[v]
\end{align}

Mathematically, if the input light curve is shifted by some integer number of days, then this
function outputs a value for $\mu_{t_0}$ that is incremented by the same number of days as desired.
We call this layer a ``time-indexing layer''. By combining the time-indexing layer with
the global max-pooling layer, we produce an encoder that is mathematically invariant to
a translation in the input of any integer number of days with appropriate transformations for
all of the latent variables. Note that we do not force the reference time to correspond
with any specific feature of the light curve other than having a weak prior on the time
of maximum light. We allow the model to automatically determine a relevant definition
of reference time on its own.

\subsection{Marginalizing Over the Amplitude} \label{sec:marginalizingamplitude}

With the time-indexing layer the model is able to converge and give reasonable predictions.
However, we find empirically that it is challenging for the model to predict the amplitude $A$.
The amplitude corresponds to an overall scaling of the model, and is degenerate with a global
scale of all of the outputs. We find that we can significantly improve the convergence of the model
by modifying the architecture to marginalize over the amplitude rather than have it as an explicit
parameter. To do this, note that given all of the other latent parameters $\bs'$, we can analytically
evaluate the conditional posterior distribution of $A$ as:
\begin{align}
    p(A|\bs',\by) = \frac{p(\by|\bs',A)p(A|\bs')}{p(\by|\bs')}
\end{align}

Here $p(\by|\bs')$ is a constant that does not depend on $A$. Assuming a Jeffreys (flat) prior
for $p(A|\bs')$, this becomes:
\begin{align}
    p(A|\bs',\by) & \propto p(\by|\bs',A) \\
    & \propto \prod \exp\left(-\frac{(\by - A \cdot d_\btheta(\bs', A=1))^2}{2 \sigma_\by^2}\right) \\
    & = \frac{1}{\sqrt{2 \pi \sigma_A^2}} \exp\left(-\frac{(A - \mu_A)^2}{2 \sigma_A^2}\right)
    \label{eq:amplitude_posterior}
\end{align}
where:
\begin{align}
    \sigma_A^2 &= \frac{1}{\sum d_\btheta(\bs', A=1)^2 / \sigma_\by^2} \\
    \mu_A &= \frac{\sum \by \cdot d_\btheta(\bs', A=1) / \sigma_\by^2}{\sum d_\btheta(\bs', A=1)^2 / \sigma_\by^2}
\end{align}

We can then evaluate the log-likelihood $\log p_\btheta(\by|\bs')$ necessary for the VAE loss function
in Equation~\ref{eq:elbo} as:
\begin{align}
    \log p_\btheta(\by|\bs') &= \int dA \cdot p(\by|\bs',A) \cdot p(A|\bs') \\
                       &= \int dA \cdot p(\by|\bs',A) + C
\end{align}
where $C$ is some arbitrary normalization constant that can be ignored because it will not affect the
minimization. We estimate this integral numerically using importance sampling with $A$ drawn from its posterior
distribution shown in Equation~\ref{eq:amplitude_posterior}. As is typically done when training
deep learning models, while training we use a single sample for our numerical estimate of the integral.

With this procedure, we analytically marginalize over the amplitude $A$ and it is no longer an explicit
parameter of our model. The Jeffreys prior that we impose on $p(A|\bs')$ is improper, which
means that we are not learning a generative model for the amplitude. This is not a problem for most
science applications as discussed in Section~\ref{sec:priors}, and a similar approach
is taken for most generative models of SNe~Ia \citep[e.g. SALT2;][]{guy07}.
We find that this model that marginalizes over the amplitude is much easier to train than a model
where the amplitude is an explicit parameter.

\subsection{Regularization of the Spectra} \label{sec:spectra_regularization}

As discussed in Section~\ref{sec:vae_symmetries}, we are training our model to predict the full
spectra of transients. We are only training using photometry in this work, which means that we
are effectively performing a deconvolution. As a result, the high-frequency components of the spectra will
not be well-constrained. To prevent the model from adding high-frequency noise to the spectra, we
add a regularization term to our loss function following a procedure similar to the one used by
\citet{crenshaw20} for estimating galaxy spectra from photometry. Given the set of wavelength bins
with index $n$, we apply a penalty on the flux difference between adjacent bins given by:
\begin{align}
    \eta \sum_i \left( \frac{d_{\text{int},\btheta,n}(\bt,\bs_i) - d_{\text{int},\btheta,n+1}(\bt,\bs_i)}
                            {d_{\text{int},\btheta,n}(\bt,\bs_i) + d_{\text{int},\btheta,n+1}(\bt,\bs_i)} \right)^2
\end{align}

Here $\eta$ is a tunable parameter that can be adjusted to determine the strength of the
regularization. We apply this penalty to the full spectra that are predicted at every time where
we have an observation. We find empirically that a value of
$\eta = 0.001$ results in a reasonable balance between reducing noise and oversmoothing the spectrum.
We find that this regularization term contributes only $\sim$0.01\% to the loss function at the
end of training for both of the datasets that we consider.

\subsection{Augmentation} \label{sec:augmentation}

As shown in \citet{boone19}, augmentation can be used to generate a wide range of different light
curves from observations of a single transient. When training our model, we apply a series of
different random transformations to our data. These transformations are all applied independently,
and we choose different transformations in each epoch.
\begin{itemize}
    \item Shift the observations by a constant offset sampled from a Normal[0, 20 days] distribution.
    \item Scale the observations by a constant factor sampled from a Lognormal[0, 0.5] distribution.
	\item Drop observations randomly, with the fraction to drop sampled from a Uniform[0, 0.5] distribution.
	\item With probability 50\%, add noise to the observations with a standard deviation sampled from a Lognormal[-4, 1] distribution.
\end{itemize}

These transformations produce a wide range of light curves, with some resembling the original light
curves and others at much lower signal-to-noises with significantly fewer observations. In contrast
to \citet{boone19}, we do not attempt to augment our light curves in redshift as that cannot be
done without making assumptions about the shape of the spectrum. We also do not attempt to reproduce
the observational properties of the instrument. Instead, we simply attempt to cover a wide range of
signal-to-noises ranging from the original values to signal-to-noises well-below the
detection threshold.

\subsection{The ParSNIP model} \label{sec:parsnip_model}

\begin{figure}
\epsscale{0.85}
\plotone{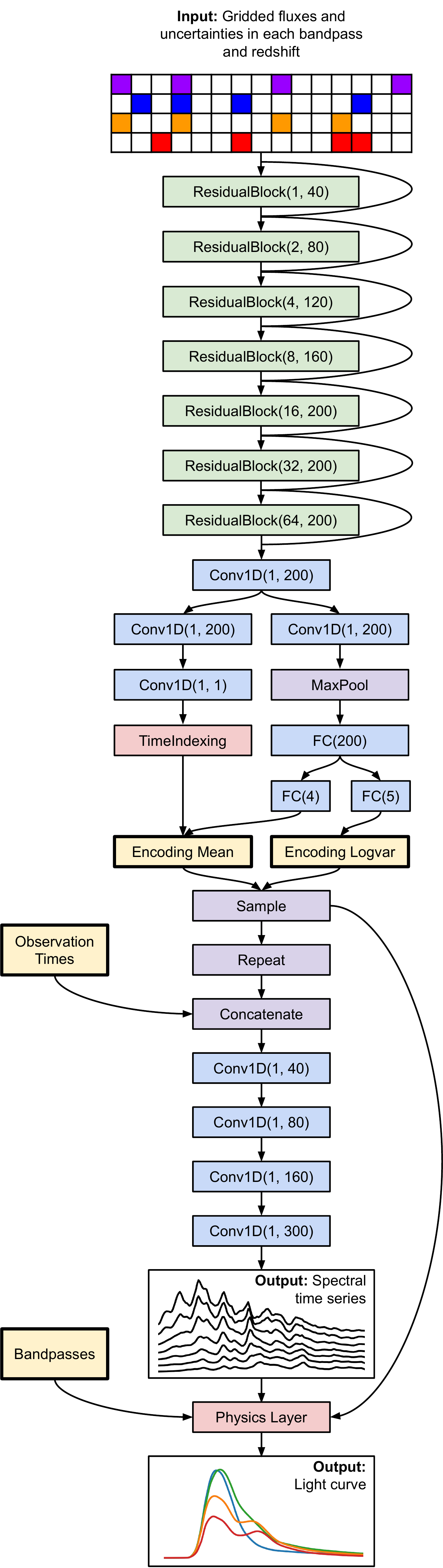}
\caption{
    Architecture of the ParSNIP model.
}
\label{fig:parsnip_architecture}
\end{figure}

We combine all of the different elements described in this Section to build a hybrid physics-VAE
model that can describe transient light curves. We call this model ``ParSNIP'' (Parametrization
of SuperNova Intrinsic Properties). A schematic of the full ParSNIP architecture is shown in
Figure~\ref{fig:parsnip_architecture}.

We implemented this model in PyTorch \citep{pytorch}. We fit the model separately
to the PS1 and PLAsTiCC datasets of transient light curves described in Section~\ref{sec:dataset}
using the Adam optimizer \citep{kingma15} with a learning rate of $10^{-3}$ and batch sizes of $128$.
We use the \texttt{ReduceLROnPlateau} feature in
PyTorch to scale the learning rate by a factor of $0.5$ whenever the optimizer fails to
reduce the loss function after 10 epochs, and we train until the learning rate is decreased to
below $10^{-5}$. This training procedure takes approximately 300 epochs. The computation
times and requirements for both training and inference are discussed in
Section~\ref{sec:computational_requirements}.

\subsection{Hyperparameters}
\label{sec:hyperparameters}

Our model has many different hyperparameters that can be tuned, including the architecture of the
network and the parameters of the optimizer. We trained 56 models on the PS1 dataset using a wide
range of different hyperparameter values. We trained on 90\% of the dataset, and kept 10\% back
for validation. We find that the model is not highly sensitive to most hyperparameters other than
the learning rate, with the loss function typically varying by $<5\%$. The results are consistent
for the training and validation sets. When the model performs similarly well on two different
configurations, we choose the one that is more computationally efficient. We show the range of
hyperparameter values that we
considered along with the optimal values from our tests in Table~\ref{tab:hyperparameters}.

\begin{deluxetable*}{cc}
\tablecaption{Hyperparameter optimization for the ParSNIP model. We show the range of values
considered for each hyperparameter along with the value that resulted in the model with the
smallest loss function in bold. When arrays are given, each element in the array refers to
the hyperparameter value for a different block in the architecture. The optimal configuration
is shown in Figure~\ref{fig:parsnip_architecture}.}
\label{tab:hyperparameters}
\tablehead{\colhead{Hyperparameter} & \colhead{Value}}
\startdata
Batch size & 16, 32, 64, \textbf{128}, 256, 512\\
Learning rate & $10^{-4}$, $2\times10^{-4}$, $5\times10^{-4}$, $\bm{10^{-3}}$, $2\times10^{-3}$, $5\times10^{-3}$, $10^{-2}$ \\
Scheduler factor & 0.1, 0.2, \textbf{0.5} \\
Encoder convolutional block types & Conv1d, \textbf{ResidualBlock} \\
Encoder convolutional block widths & [20, 40, 60, 80, 100, 100, 100] \\
& \textbf{[40, 80, 120, 160, 200, 200, 200]} \\
& [32, 64, 128, 128, 128, 128, 128] \\
& [32, 64, 128, 256, 256, 256, 256] \\
& [32, 64, 128, 256, 512, 512, 512] \\
Encoder fully connected block sizes & 100, 128, \textbf{200}, 256, 400, 512 \\
Decoder architecture & [100] \\
& [50, 100] \\
& [100, 200] \\
& \textbf{[40, 80, 160]} \\
& [64, 128, 256] \\
& [32, 64, 128, 256] \\
\enddata
\end{deluxetable*}

\subsection{Dimensionality of the Intrinsic Latent Representation}

\begin{figure}
\plotone{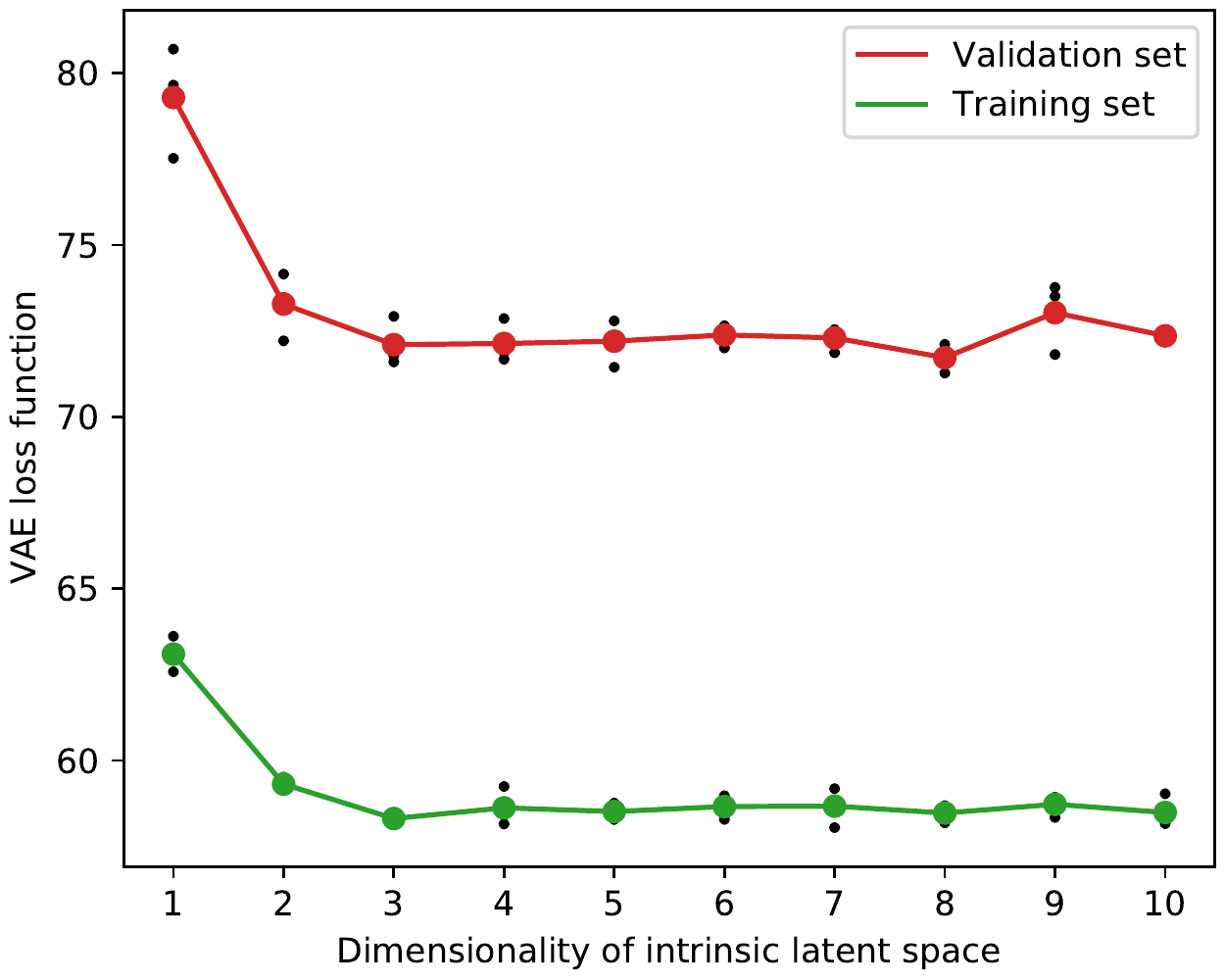}
\caption{
    VAE loss function for different dimensionalities of the intrinsic latent representation.
    For each number of dimensions we train three models (shown as black dots), and
    we shown the mean value of the loss function across all three models with
    colored markers. Beyond three dimensions we see little improvement in the VAE loss function.
}
\label{fig:vae_dimensionality}
\end{figure}

Another important hyperparameter is the dimensionality of the intrinsic
latent representation. We trained the ParSNIP model with varying numbers of intrinsic
dimensions and with all of the other hyperparameters kept at their optimal
values shown in Table~\ref{tab:hyperparameters}. For each model, we evaluated
the VAE loss function shown in Equation~\ref{eq:elbo} on both the training
and validation subsets of the PS1 dataset. The results of this procedure
are shown in Figure~\ref{fig:vae_dimensionality}.

We find that the loss function improves when increasing the dimensionality up to
three for both the training and validation sets, but there is little benefit from
adding more dimensions beyond three. For models with large numbers
of dimensions, we find that most of the dimensions are not used by the VAE, and
it simply outputs uninformative $\mathcal{N}(\bm{0}, \bm{1})$ posteriors for
all of the transients in some dimensions. We choose to use a three-dimensional
model for the rest of our experiments.

\section{Model Performance} \label{sec:performance}

\subsection{Reproducing Light Curves}

\begin{figure*}
\epsscale{1.15}
\plotone{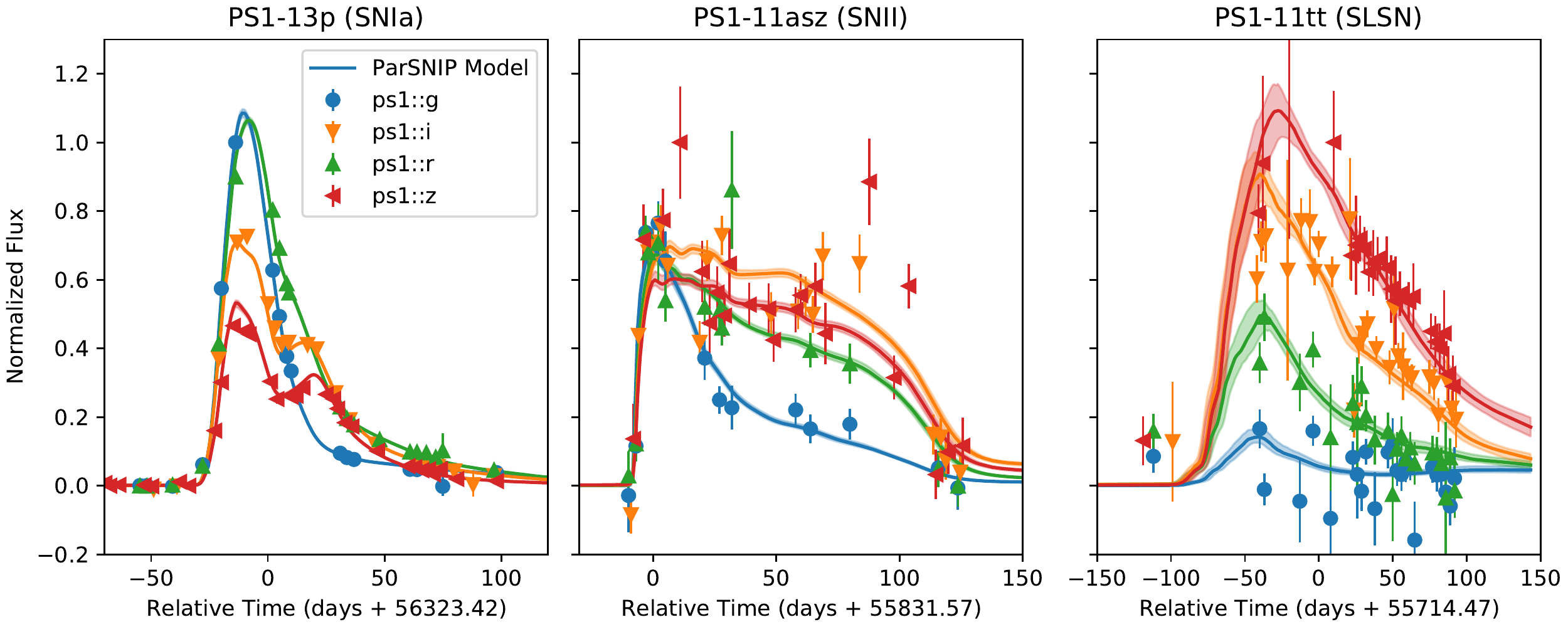}
\caption{
    Examples of out-of-sample predictions of the ParSNIP model on the validation subset of the PS1 dataset.
    These light curves were not included in the dataset that was used to train the model.
    Each panel shows the predictions for a different light curve. The observations are
    shown as individual points with their uncertainties, and the different colors correspond
    to the different bands as shown in the legend. The mean model predictions are shown as
    a solid line, and the 1-sigma uncertainties from varying the VAE latent parameters
    are shown with shaded contours.
}
\label{fig:model_examples}
\end{figure*}

To study how well our model can reproduce light curves, we trained the ParSNIP
model on 90\% of the light curves in the PS1 dataset
described in Section~\ref{sec:dataset} with a randomly selected
10\% of the light curves held out for validation. We show examples of the
models for different light curves in the validation subset in
Figure~\ref{fig:model_examples}. We find that the ParSNIP model is able to
generalize well on the PS1 dataset with accurate models for the vast majority
of transients.

We calculated the model residuals for all of the observations in our dataset.
For the PS1 training set, we find that the model residuals are consistent with the
statistical uncertainties which suggests that may we have somewhat
overfit the training set. For the validation set, we find
that the distribution of the residuals has a tight core with a dispersion of $\sim$0.06~mag
when statistical uncertainties are taken into account.
We do however see large residuals for some light curves, particularly ones of
transient types that are not well represented in the training set. For observations
with a statistical uncertainty of less that 0.05~mag, we find
that 94\% of observations have residuals of less than 0.2~mag, and $99.2$\% of observations
have residuals of less than 0.5~mag.

We performed the same comparison for a version of the ParSNIP model trained
on 90\% of the light curves in the PLAsTiCC dataset. In this case, we find that the model residuals have a
tight core with a dispersion of $\sim$0.04~mag
when statistical uncertainties are taken into account.
For observations with a statistical uncertainty of less than 0.05~mag, we find
that 97\% of observations have residuals of less
than 0.2~mag and $99.7$\% of observations have residuals of less than 0.5~mag,
with no major differences between the training and validation sets. The PLAsTiCC dataset was
simulated and the subset that we used to train the model contains more than 100 times as
many transients as the PS1 dataset which likely explains the difference in training/validation
performance between the PS1 and PLAsTiCC datasets.
Overall, we find that the ParSNIP model is able to produce good models for the vast majority of
the light curves in both datasets.

\subsection{The ParSNIP Intrinsic Latent Space}

\begin{figure*}
\epsscale{1.15}
\plottwo{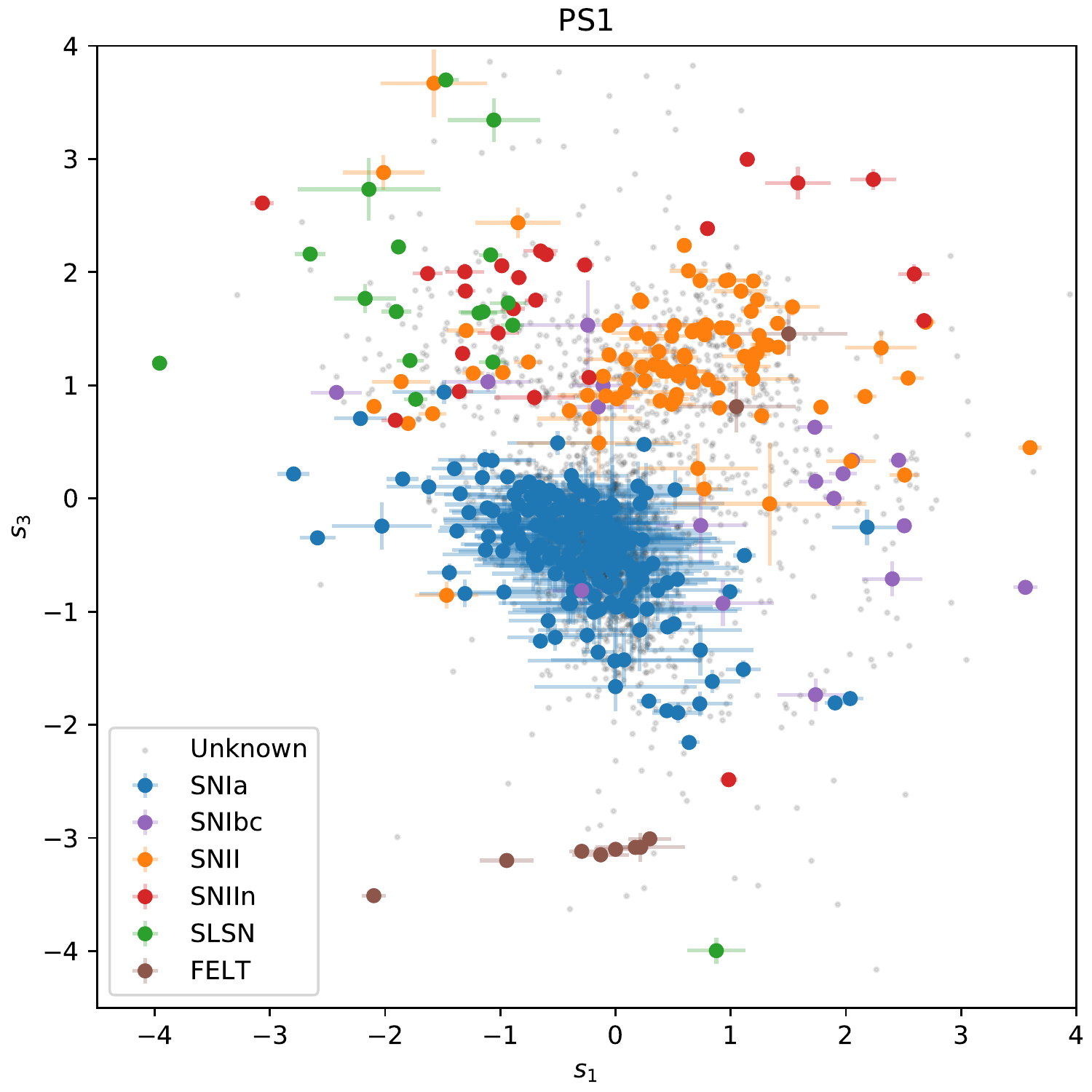}{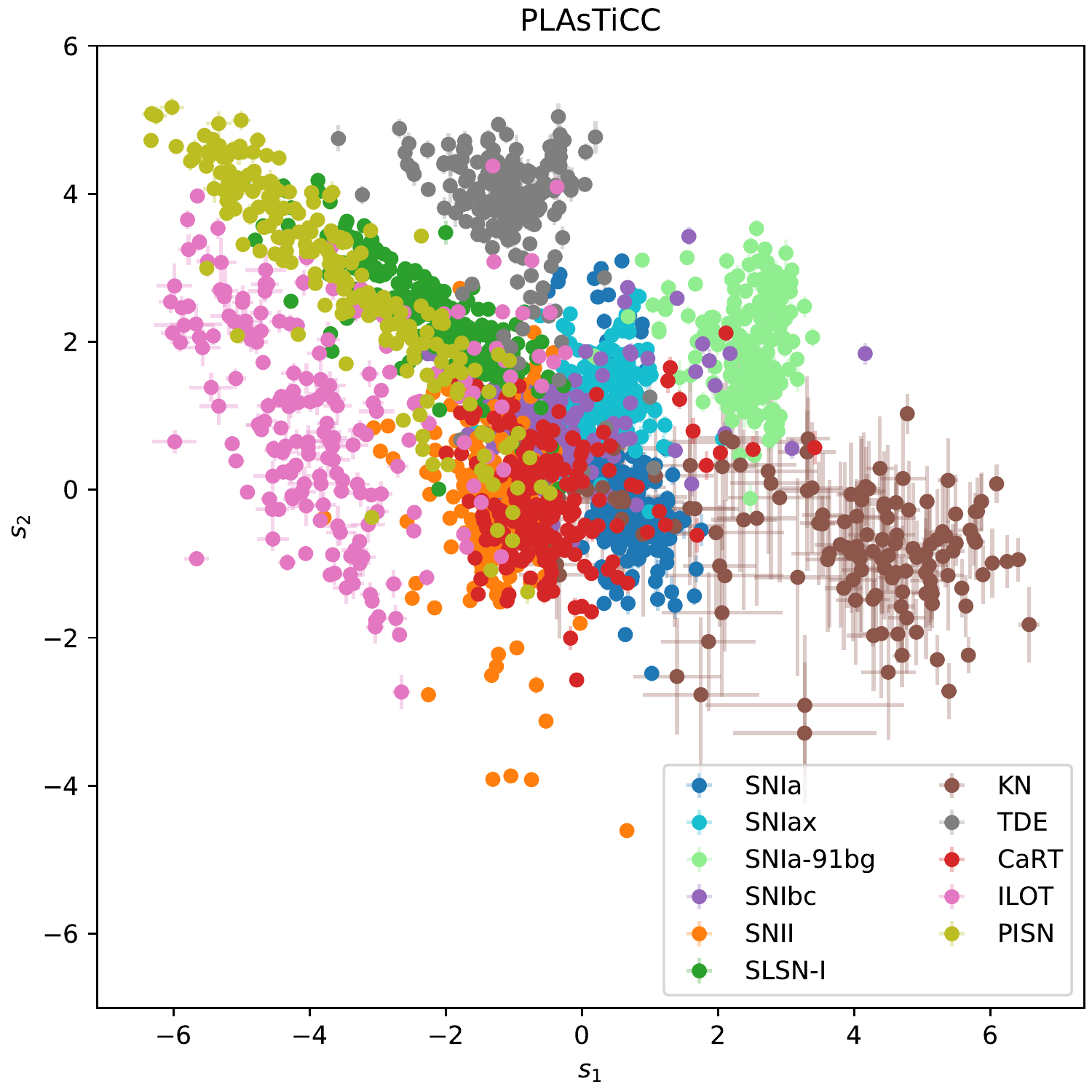}
\caption{
    Visualization of the intrinsic latent spaces learned when trained on the PS1 (left panel) and PLAsTiCC
    (right panel) datasets. We show two of the $\bs_i$ latent variables for each transient in the dataset,
    and we color each transient by its type if known. We limit this plot to show a maximum of 200 transients
    for each type to highlight
    rarer types. Despite the fact that the types were not used when training the ParSNIP model,
    the model groups transients of the same type at similar locations in the latent space. Note
    that these plots show only two dimensions of the intrinsic latent space:
    each light curve also has measurements of another $\bs_i$ latent variable, color, and amplitude.
}
\label{fig:representations}
\end{figure*}

As described in Section~\ref{sec:vae_symmetries}, the ParSNIP model learns a
three dimensional intrinsic latent space $\bs_i$ describing the intrinsic diversity of transients.
To study the structure of the learned latent spaces, we look at where transients with known labels
are located in the latent space for both of our datasets. The results of this procedure
are shown in Figure~\ref{fig:representations}. Most of the different transient types
are well separated despite the fact that the type labels were not used when
training the ParSNIP model.

\begin{figure*}
\plottwo{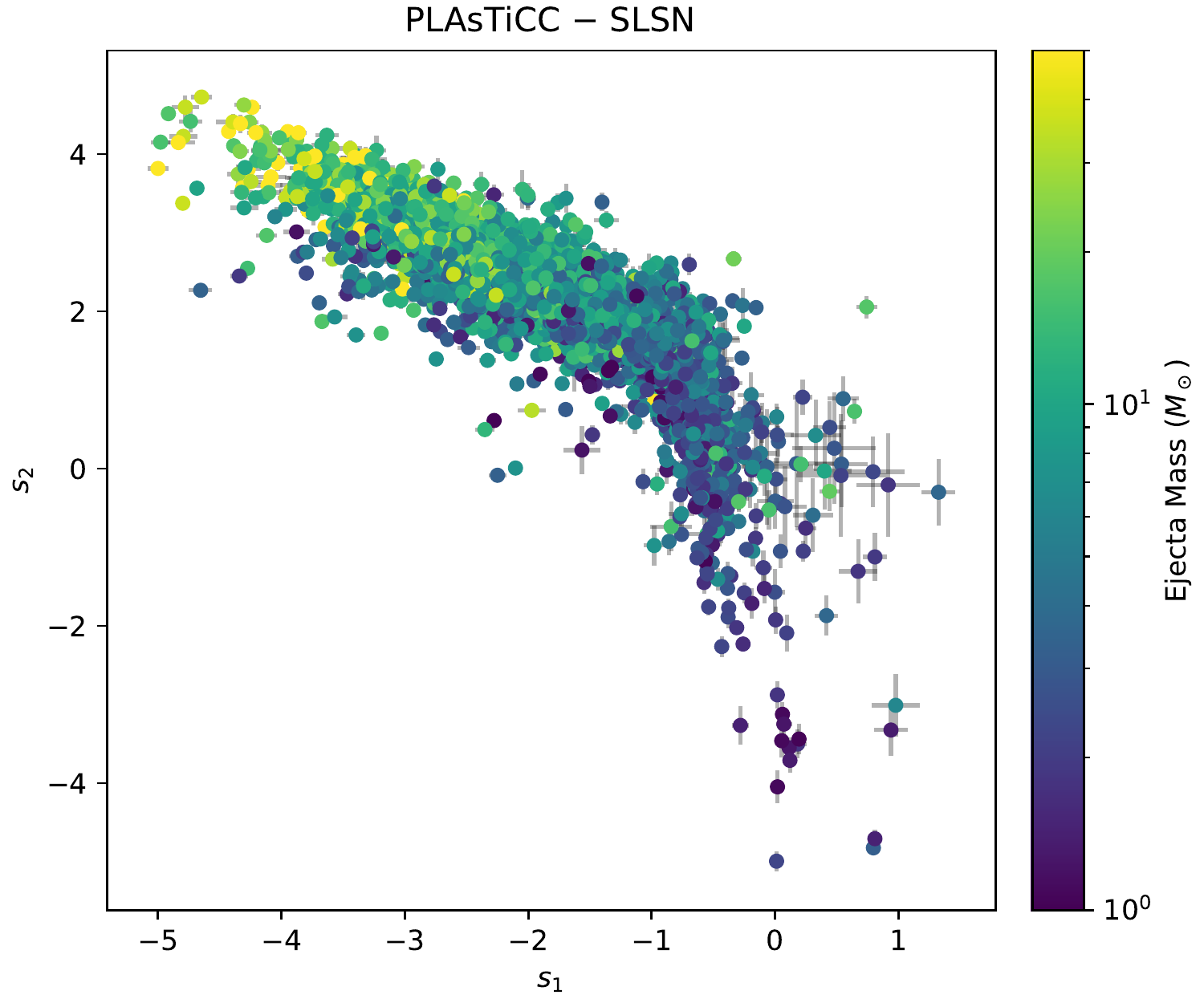}{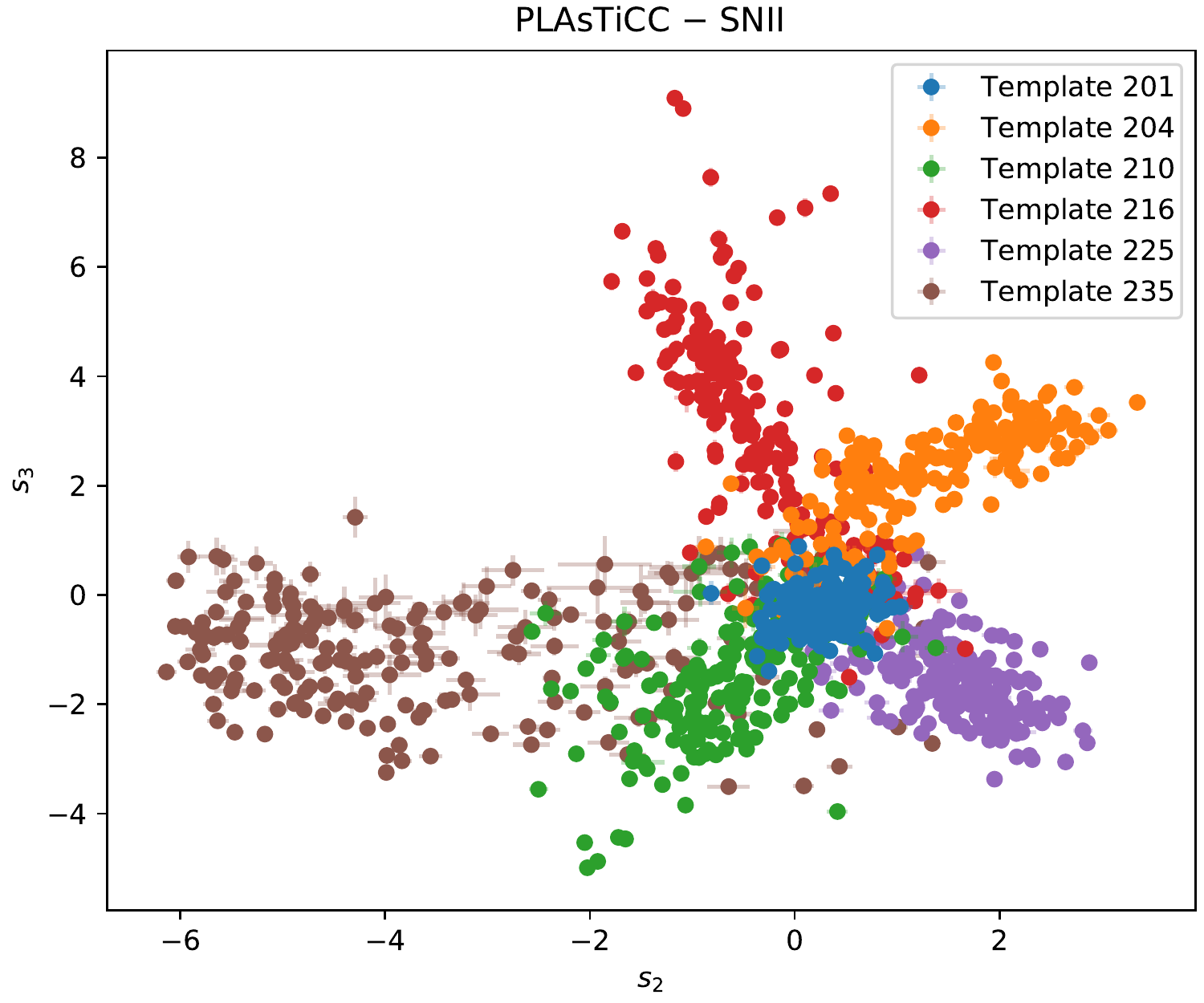}
\caption{
    Examples of diversity of the ParSNIP intrinsic representations for transients of the same type
    in the PLAsTiCC dataset. Left panel: The variation of ParSNIP $s_1$ and $s_2$ latent variables for superluminous
    supernovae is explained by differences in the ejecta mass parameter (shown in color) that went
    into the simulations. Right panel: Some of the Type~II supernovae in the PLAsTiCC dataset
    were simulated using discrete templates. We show the $s_2$ and $s_3$ latent variables for light curves
    simulated from six of these templates, and find that the light curves simulated from each template
    cluster at similar locations in the ParSNIP intrinsic latent space.
}
\label{fig:submodel_representations}
\end{figure*}

For transients of the same type, we find that differences in the ParSNIP intrinsic latent
variables $\bs_i$ correspond to differences in the intrinsic properties
of the transients. This can easily be see using the PLAsTiCC
dataset where all of the parameters that were used for simulating each
light curve are known. In the left panel of Figure~\ref{fig:submodel_representations},
we show that the ParSNIP latent space of superluminous supernovae is capturing the
variation in ejecta mass that went into the simulations. The SNe~II were simulated
using a small number of discrete templates, and we show where light curves simulated
from some of these templates are located in the latent space in the right panel of
Figure~\ref{fig:submodel_representations}. The ParSNIP model identifies these discrete
templates and clusters light curves simulated from each of them in the latent space. 

As discussed in Section~\ref{sec:methods}, the ParSNIP model was constructed to produce
an intrinsic latent space that is invariant to how the light curve was observed.
As expected, we find that well-measured light curves simulated with similar parameters
are embedded at very similar locations in the intrinsic latent space even when those light
curves are observed at different redshifts, with different brightnesses, or with
different amounts of host galaxy dust. There are no major correlations between any
of these properties of the observations and the intrinsic latent variables $\bs_i$.

\begin{figure*}
\epsscale{1.15}
\plotone{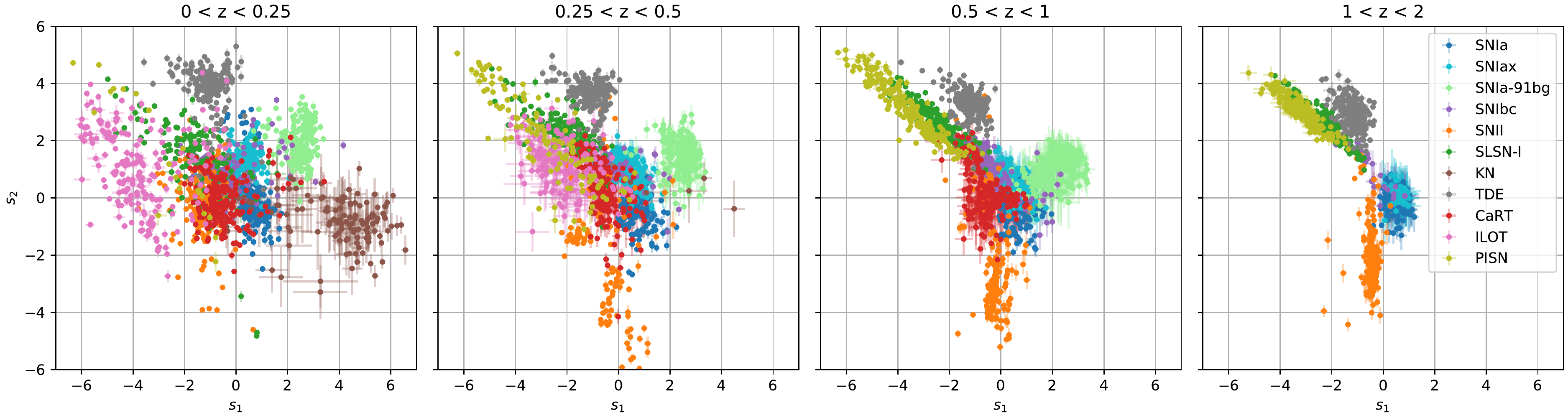}
\caption{
    Visualization of the intrinsic latent space learned for the PLAsTiCC dataset in
    different redshift
    ranges. The redshift range of each panel is shown above the panel. In each panel, we show the
    $s_1$ and $s_2$ latent variables for 200 transients of each type, and we color each point by its
    type. The ParSNIP model was constructed to produce an intrinsic latent space that is invariant
    to observing conditions such as redshift, and we find that the latent spaces are nearly
    identical in each of the different redshift slices. The slight differences such as the
    narrowing of the SNe II distribution at high redshifts are due to Malmquist bias rather than
    a limitation of the ParSNIP model. Note that some transient types have a limited redshift
    range in the PLAsTiCC simulations so they do not appear in some of the redshift slices.
}
\label{fig:plasticc_representation_z}
\end{figure*}

To demonstrate the invariance of the ParSNIP model to redshift, we show the recovered intrinsic
latent space for the PLAsTiCC dataset in different redshift slices in Figure~\ref{fig:plasticc_representation_z}.
We find that the recovered latent space is almost identical in all of the different redshift
slices. There are minor differences
across redshift bins for some of the transient types. For example, the SNe~II have a much broader
distribution in the lower redshift bins compared to the higher redshift
bins. These differences are due to Malmquist bias rather than a limitation of the ParSNIP model.
At low redshifts nearly all SNe~II are detected, but at high redshifts only the SNe~II that
are intrinsically very bright are detected. For the PLAsTiCC dataset, this means that at low
redshifts the SNe~II come from a wide range of templates giving a wide distribution in latent
space, but at high redshifts the majority of the observed SNe~II come from a single template
(number 235) and are clustered tightly in the latent space.

\subsection{Comparison with SALT2} \label{sec:salt_comparison}

The SALT2 model \citep{guy07} is commonly used to model the light curves of SNe~Ia.
This model is a linear model with one parameter $x_1$
describing the variability in the observed widths of SN~Ia light curves, and a second
parameter $c$ describing the differences in color (which will capture host galaxy dust).
As described in Section~\ref{sec:vae_symmetries}, the ParSNIP model is effectively a generalization
of the SALT2 model. In this subsection, we compare the performance of both of these models
for SNe~Ia. Note that the PLAsTiCC SN~Ia light curves were simulated using SALT2. For a fair
comparison, we only consider the version of the ParSNIP model that was trained
on real PS1 light curves and is thus independent of SALT2. 

We investigated whether the latent space that the ParSNIP model learns
for SNe~Ia captures the $x_1$ and $c$ parameters of the SALT2 model.
To do this, we fit the SALT2 model to all of the light curves that were labeled as SNe~Ia 
in the PS1 dataset and compared the results with the ParSNIP model on the same subset of light curves.
We limited our comparison to light curves that are well-fit by SALT2, which we define as having
a fit that converges, a best-fit SALT2 $x_1$ parameter between $-5$ and $+5$, an uncertainty on $x_1$ of
less than 0.5, an uncertainty on SALT2 $c$ of less than 0.1, and at least one observation before
maximum light. 

\begin{figure*}
\plottwo{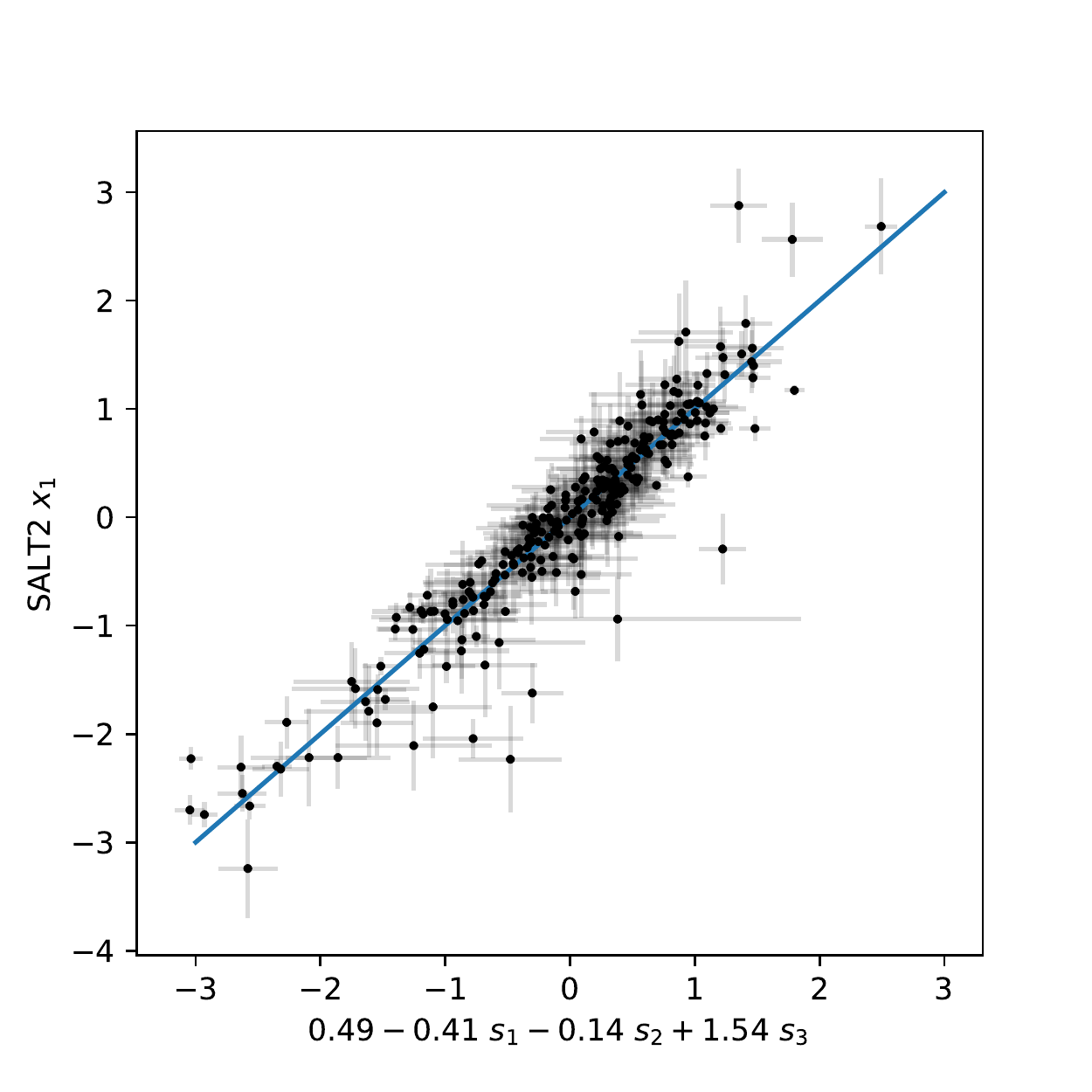}{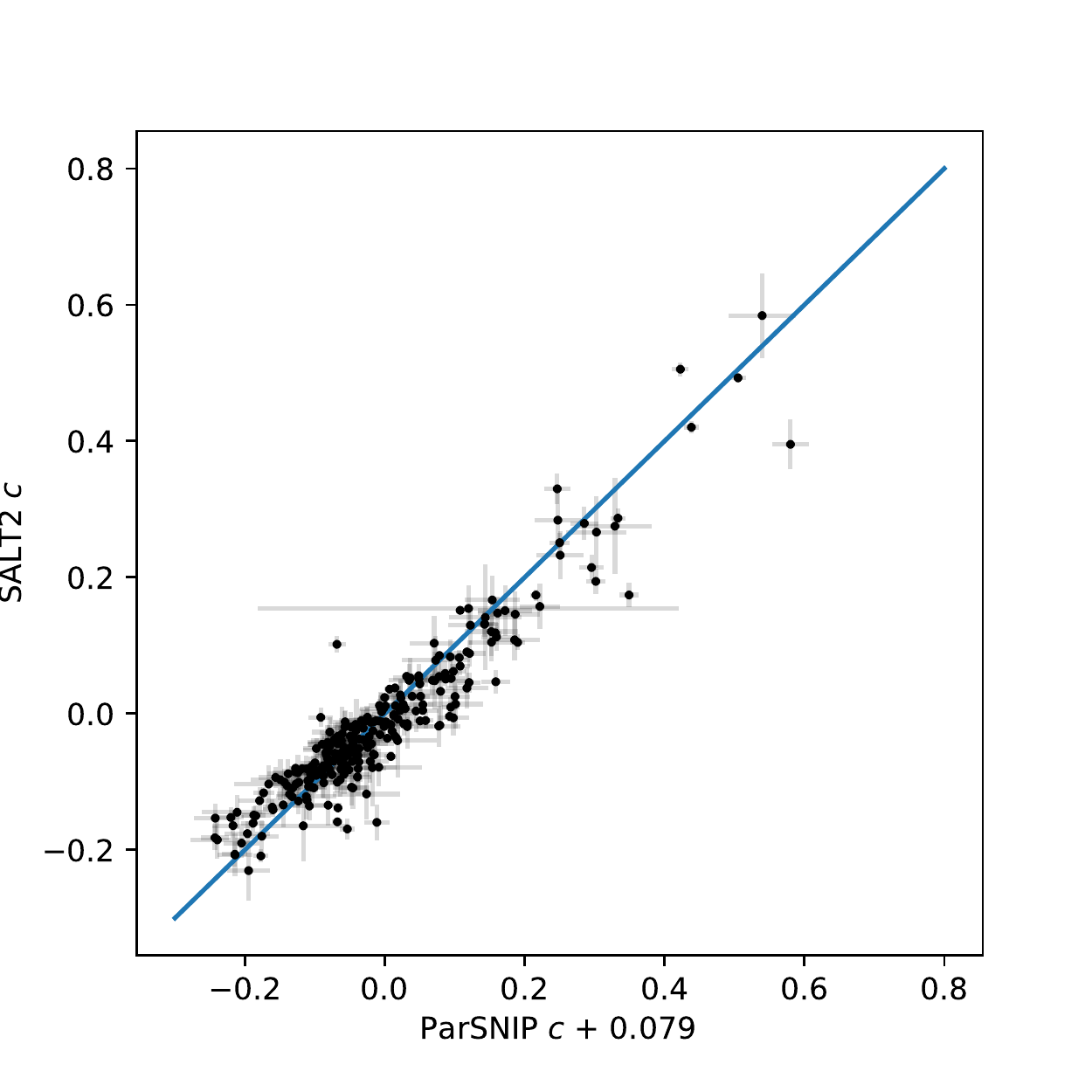}
\caption{
    Reproducing the SALT2 model parameters from the ParSNIP latent space.
    Left panel: SALT2 $x_1$ as a function of the best-fit linear combination of the
    ParSNIP intrinsic latent variables $s_i$. Right panel: SALT2 color as a function of the ParSNIP color
    with a constant offset.
    The one-to-one line is shown in blue in both plots. We are able to recover both SALT2
    $x1$ and $c$ from the ParSNIP latent space with high accuracy.
}
\label{fig:salt_parameters}
\end{figure*}

The ordering of the intrinsic latent variables $\bs_i$ in the ParSNIP model is arbitrary. To effectively
compare our parameterization of the intrinsic diversity to $x_1$, we solve for the linear
combination of the ParSNIP latent variables $\bs_i$ that best models $x_1$:
\begin{align}
    x_1^\text{ParSNIP} = \gamma_0 + \sum_n \gamma_n s_{i,n}
\end{align}
We solve for the coefficients $\gamma_n$ that minimize the sum of squared differences between
$x_1^\text{ParSNIP}$ and $x_1$. Note that we do not include measurement uncertainties in
this procedure because both the SALT2 and ParSNIP models were fit to the same observations
and the uncertainties on the $\bs_i$ and $x_1$ latent variables are likely very highly correlated.
The results of this procedure are shown in the left panel of Figure~\ref{fig:salt_parameters}.

In the right panel of Figure~\ref{fig:salt_parameters}, we show the SALT2 $c$ parameter
in terms of the ParSNIP $c$ parameter. We include a constant offset in this comparison
given by the difference in the median colors between both datasets to account for
different arbitrary zeropoints.
We are able to accurately predict the SALT2 $x_1$ and $c$ parameters
from the ParSNIP latent space for all of these SNe~Ia,
with 88\% and 89\% of the variance explained respectively.
The residuals from the predictions for $x_1$ have an NMAD of 0.24 and a standard
deviation of 0.34. The differences between the ParSNIP $c$ and SALT2 $c$ values
have an NMAD of 0.037~mag and a standard deviation of 0.044~mag. The differences between
the ParSNIP $c$ and SALT2 $c$ values are highly correlated with the $s_3$ parameter
($\rho$=0.64) which simply reflects the fact that the color zeropoint is arbitrary and can
vary across the parameter space. Our prediction of SALT $c$ could be further improved
if this were taken into account. To summarize, given the location of an SN~Ia in the
ParSNIP latent space we are able to accurately predict its SALT2 parameters. This implies that
the ParSNIP latent space contains all of the information that is captured by the
SALT2 parameterization of SNe~Ia.

\begin{figure}
\plotone{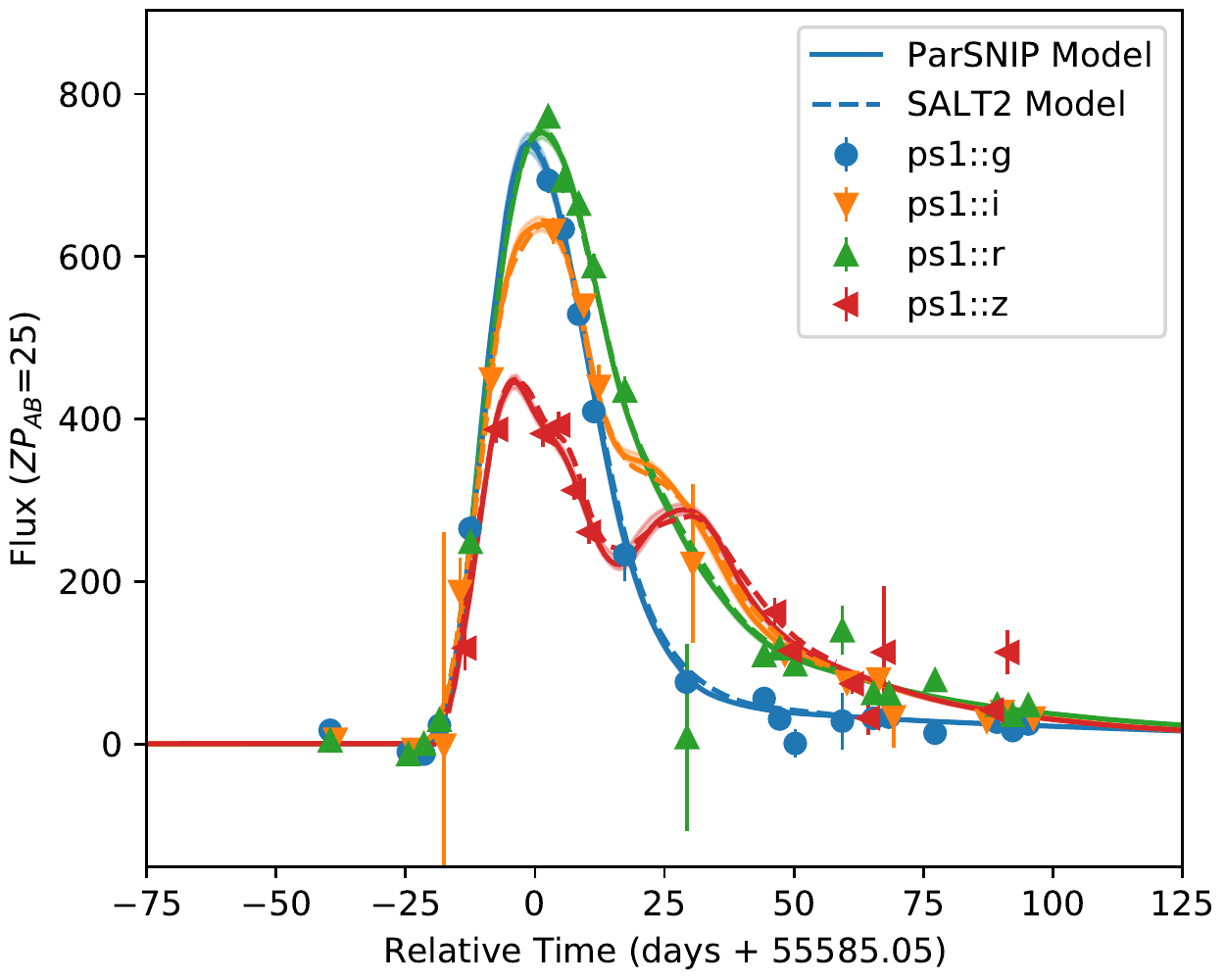} \\
\plotone{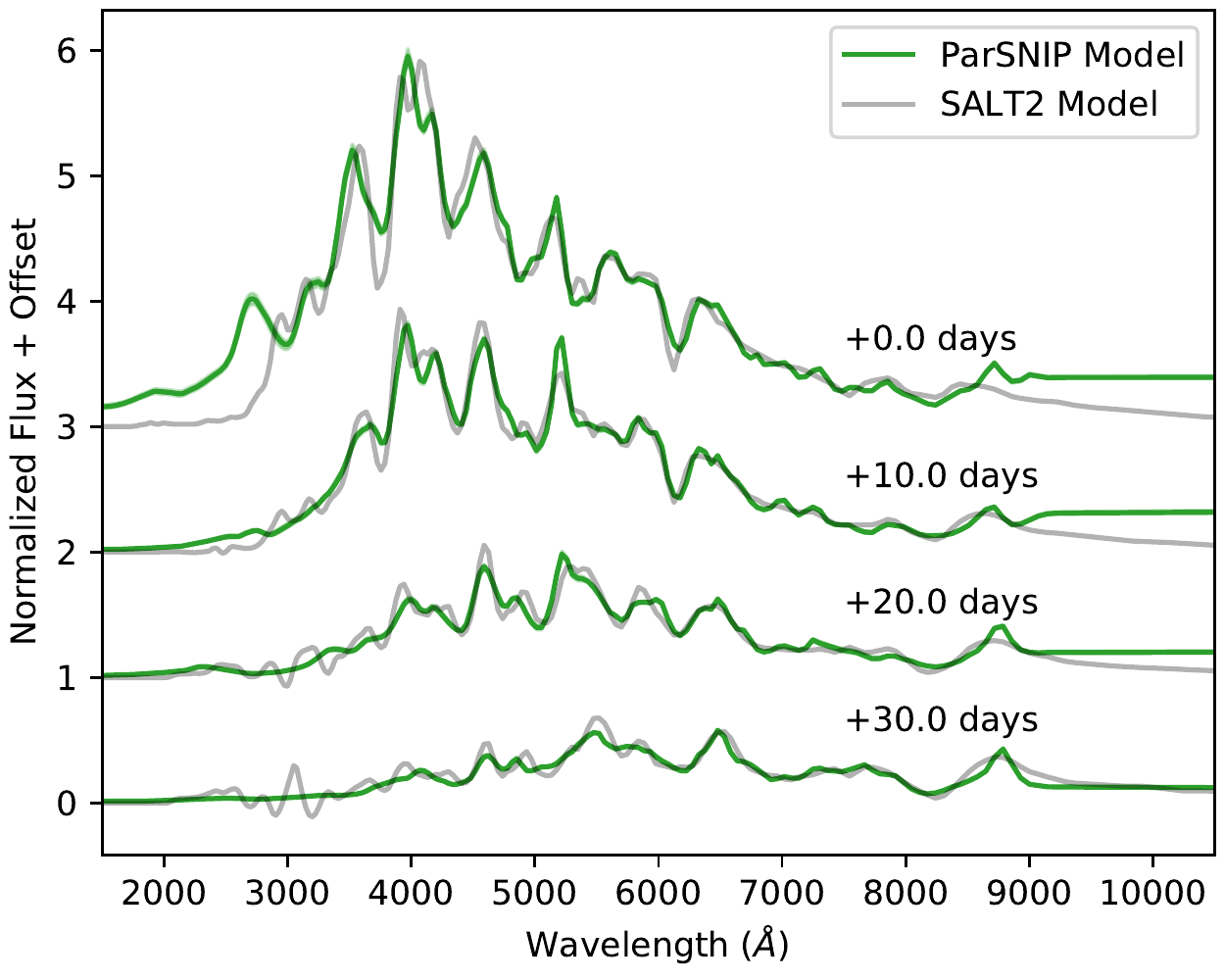}
\caption{
    Comparison of the SALT2 and ParSNIP models for the SN~Ia PS1-11bk.
    Top panel: Comparison of the light curves. The ParSNIP model is shown with solid
    lines, and the SALT2 model is shown with a dashed line. Bottom panel: Comparison
    of the spectra predicted by both models at a range of different times. We find
    good agreement between the SALT2 and ParSNIP models other than in the UV and IR
    regions that are not covered by many observations.
}
\label{fig:snia_spectrum_comparison}
\end{figure}

We compared how well the SALT2 and ParSNIP models are able to reproduce light
curves. An example of the light curve fits of both models to PS1-11bk is shown
in the top panel of Figure~\ref{fig:snia_spectrum_comparison}. The quality of the
light curve fits are comparable. For the training set (validation set), we find a
median reduced $\chi^2$ of 1.42 (1.68) for ParSNIP fits to SNe~Ia
and 1.40 (1.21) for SALT2 fits. Note that for SALT2 we found the exact maximum
likelihood values of the model parameters by fitting the model to the light curves.
For the ParSNIP model, we instead used the encoder to approximate the
maximum-a-posteriori values of the model parameters. 
All of the ParSNIP reduced $\chi^2$ values would decrease if a full fit were performed
to find the true maximum likelihood values of the model parameters. Nevertheless,
we find that the ParSNIP model is achieving comparable reconstruction performance
to the SALT2 model despite being trained on a much lower quality dataset.

Both SALT2 and ParSNIP predict the full spectral time series of the light curves
that they fit. We compare the predicted spectra for PS1-11bk at a range of different
times in the lower panel of Figure~\ref{fig:snia_spectrum_comparison}. The SALT2
model was trained using spectra of a large number of SNe~Ia and is known
to provide a good description of the spectra of SNe~Ia. In contrast,
the ParSNIP model was trained using only photometric observations. Nevertheless,
the ParSNIP model accurately reproduces all of the major spectral features
of SNe~Ia as seen in the SALT2 model. The ParSNIP model does have some non-physical
behavior at UV and IR wavelengths (as does SALT2) where the PS1 bandpasses have limited
coverage. This could be improved by including additional followup observations in
the training. We discuss this further in Section~\ref{sec:including_spectra}.
We stress that the ParSNIP model was able to learn the spectrum of an SN~Ia
by effectively deconvolving photometric observations of SNe~Ia
at a wide range of different redshifts, and no spectra were included in the training
dataset.

\subsection{Models of Other Kinds of Transients} \label{sec:models_other}

While empirical models such as SALT2 have been previously developed for SNe~Ia,
models of other kinds of transients are typically much more primitive
due to the lack of spectroscopic observations and more complex variability.
Our techniques produce a generative model for the spectral time series of all of
these transients. To evaluate the performance of these models, we compare the
spectral time series predicted by ParSNIP to observed spectra for two different
core-collapse supernovae: PS1-12cht, a Type~IIn supernova and PS1-12baa, a
peculiar Type~Ic supernova. We retrieved spectra for these two supernovae
from the Open Supernova Catalog \citep{guillochon17}. Both of these supernovae
are from classes that are not well-sampled in the PS1 dataset,
so we show results from a run where these supernovae were included in the training
of the ParSNIP model.

\begin{figure*}
\plottwo{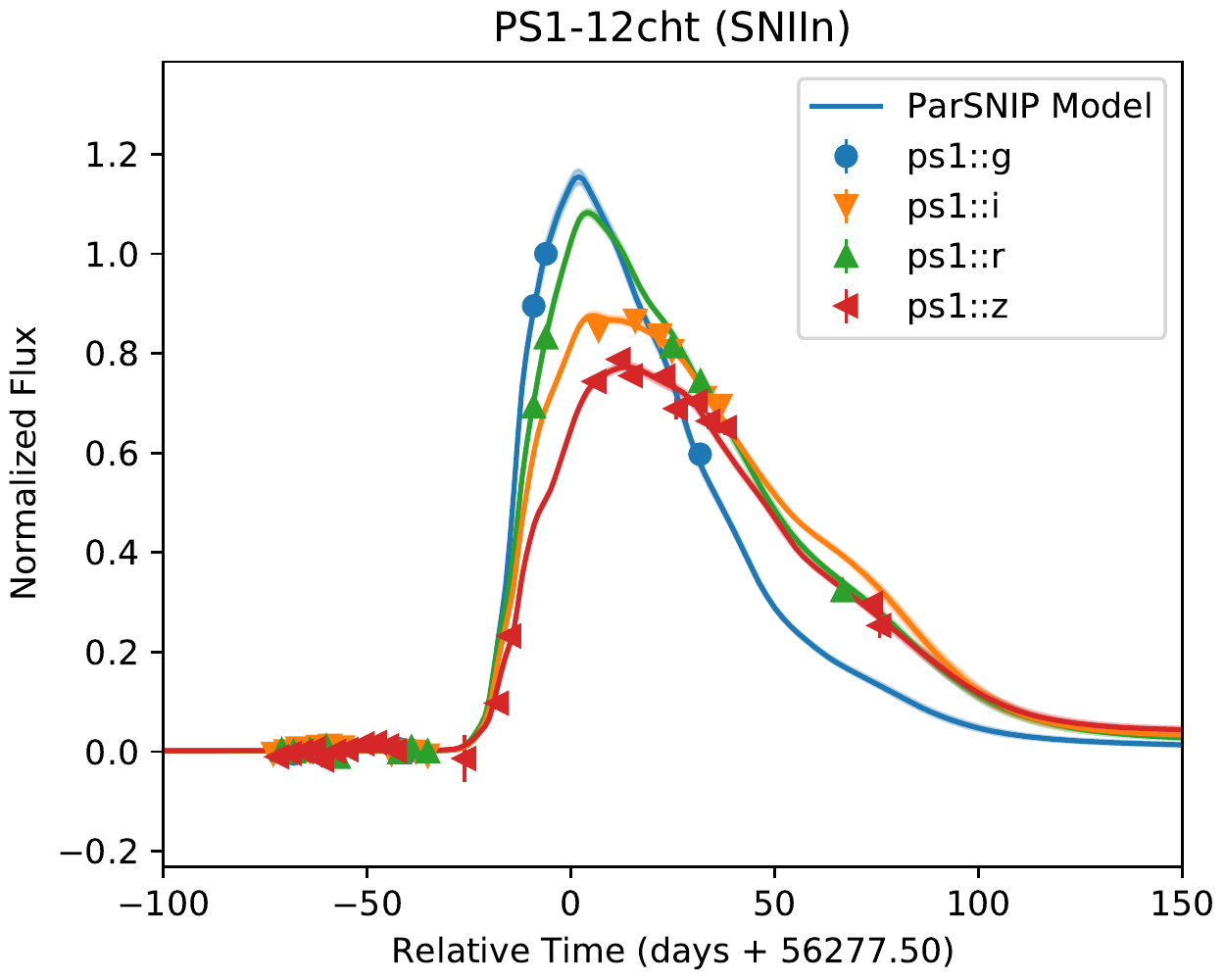}{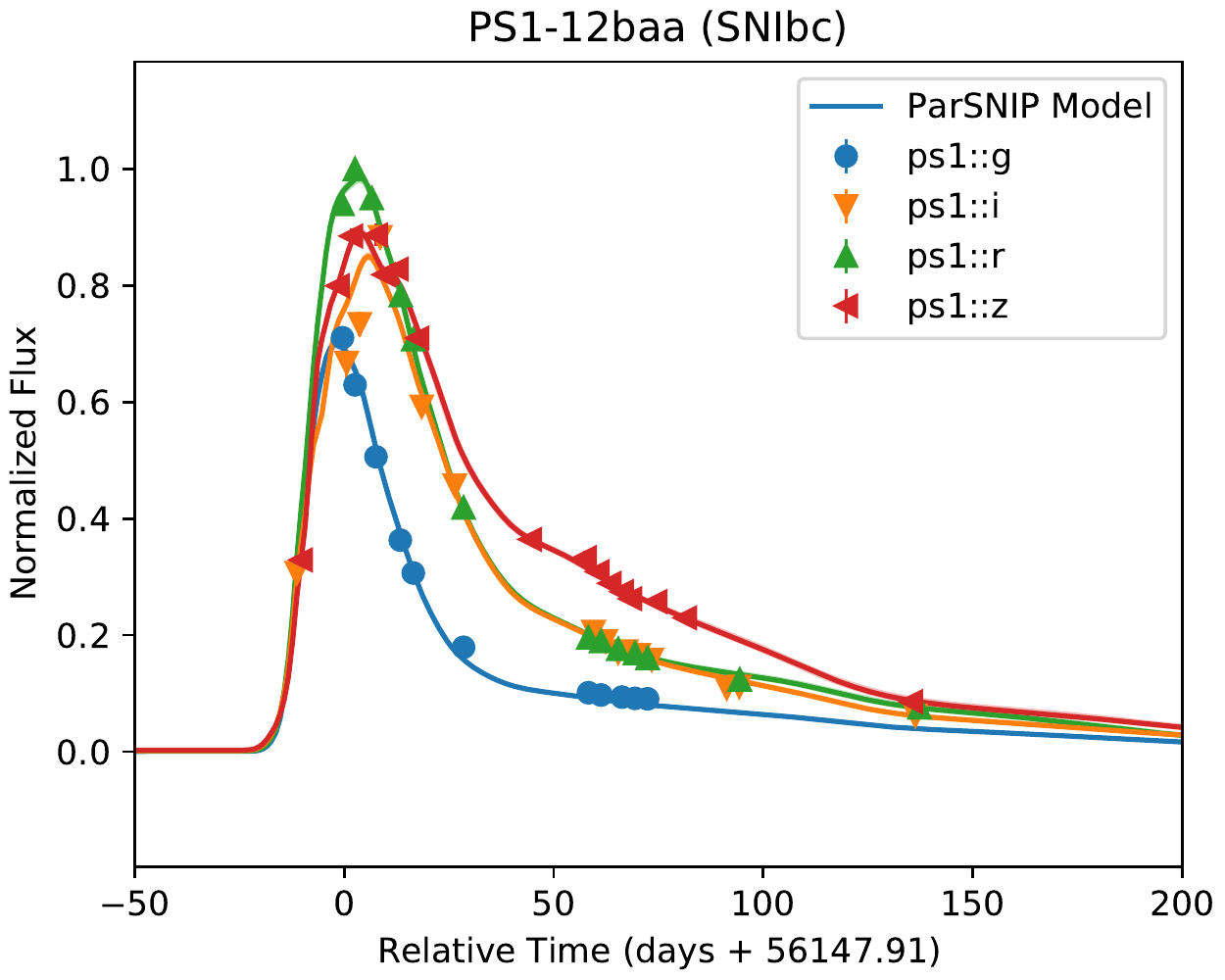} \\
\plottwo{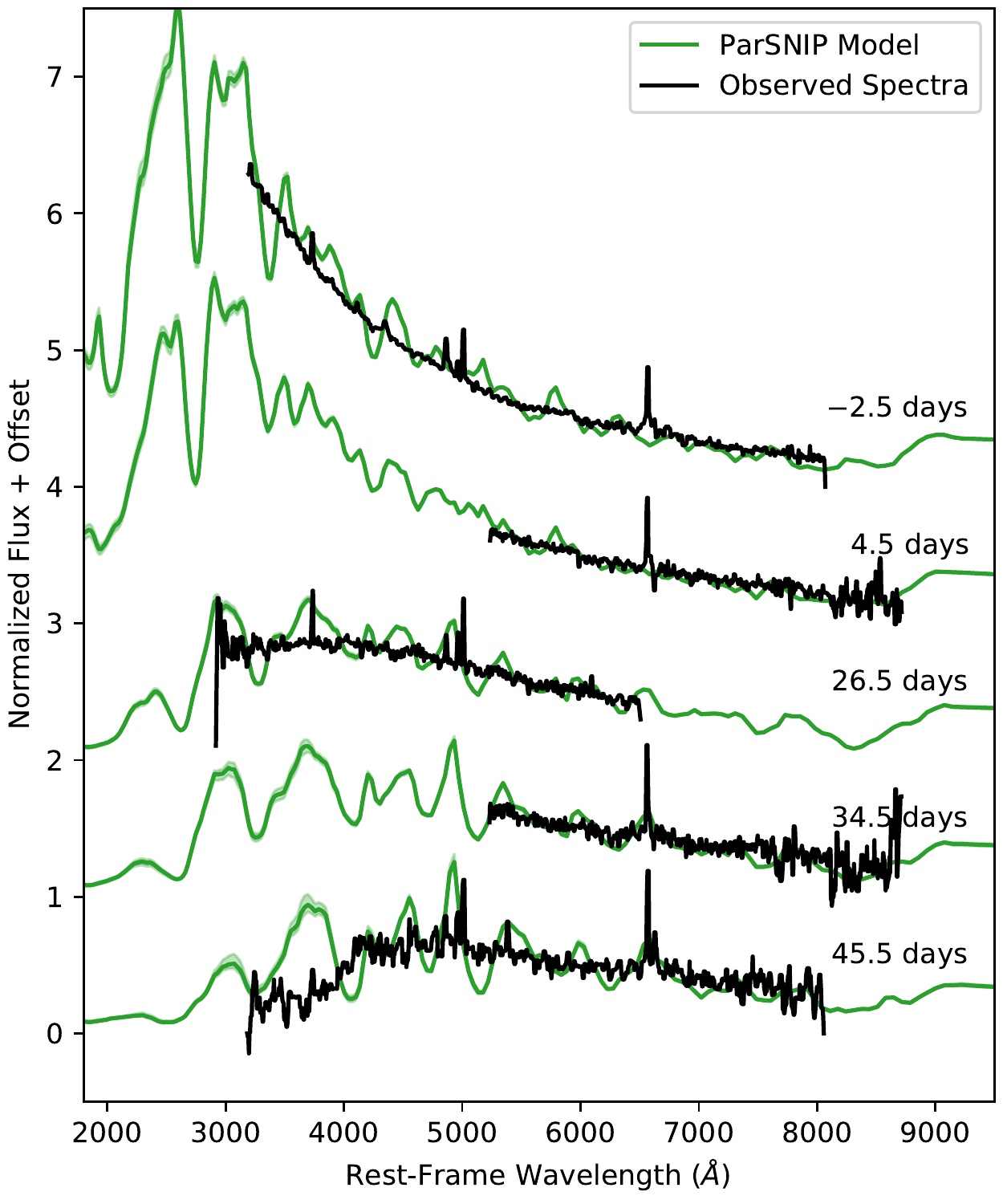}{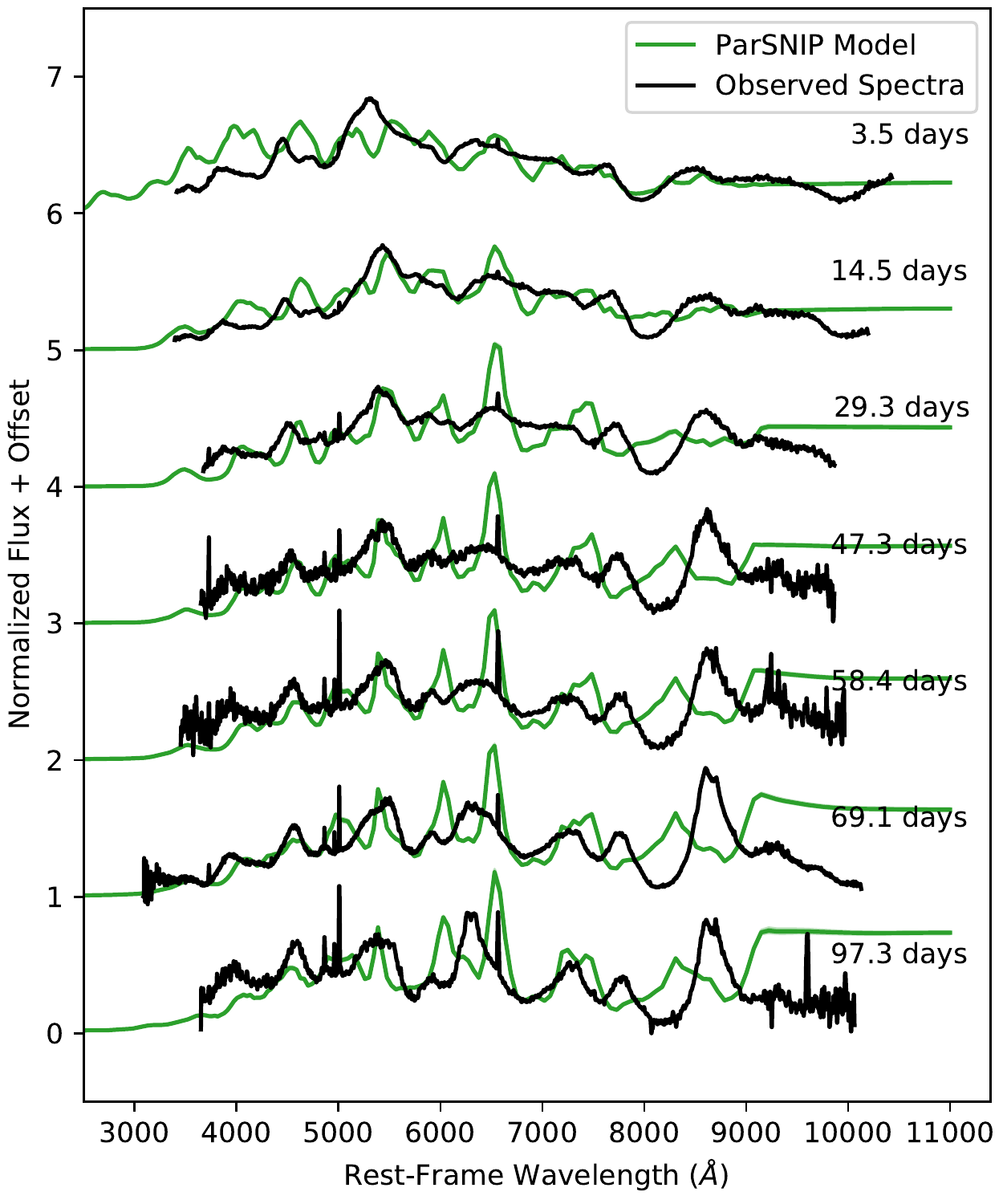}
\caption{
    Comparison of the ParSNIP model to observed spectra of core-collapse supernovae.
    The left two panels show the model and data for PS1-12cht, a Type~IIn supernova,
    and the right two panels show the model and data for PS1-12baa, a peculiar Type~Ic
    supernova. The top two panels show the observed PS1 photometry and the ParSNIP
    model fit to that photometry. The bottom two panels show observed spectra
    for each of these supernovae from PESSTO \citep{smartt15} and
    LOSS \citep{shivvers19} respectively along with spectra from the ParSNIP model evaluated
    at the same phases. We normalize all of the spectra to the flux at 6000\AA.
    We find that the broad structure of the ParSNIP predicted spectra agree well
    with the observed spectra despite the fact that the ParSNIP model was only
    trained using photometry.
}
\label{fig:other_spectrum_comparison}
\end{figure*}

In the left panel of Figure~\ref{fig:other_spectrum_comparison}, we show the
light curve and spectra of PS1-12cht, a Type~IIn supernova. These spectra were
obtained by the PESSTO collaboration \citep{smartt15} and are relatively featureless.
We find that the ParSNIP model produces good predictions of the overall shape of the spectra
and the photometry, although we do see some ringing of the spectra at bluer wavelengths.

In the right panel of Figure~\ref{fig:other_spectrum_comparison}, we show the
light curve and spectra of PS1-12baa, a peculiar Type~Ic supernova. These spectra
were obtained by the LOSS collaboration \citep{shivvers19}. The spectra of this supernova
show many different emission lines, especially at later times. The ParSNIP model
predictions agree well with the observed spectra at early times and reproduce most
of the observed spectral features. At later times, we find that the ParSNIP model
is able to predict the broad structure of the spectrum, but it struggles to predict
the exact locations and widths of all of the emission lines.

This is not surprising: the ParSNIP model
was only trained on photometry, so the estimates of the spectra come from
effectively deconvolving the photometry of many transients at different redshifts.
PS1-12baa is a peculiar Type~Ic supernova, and there are very few examples of
similar supernovae in the PS1 dataset. The deconvolved spectra are therefore not
very well constrained. This could be addressed by training on a larger dataset,
such as the one that the Rubin Observatory will produce, or by including spectra
or additional followup observations in the training process. These options are discussed in
Section~\ref{sec:including_spectra}.

\subsection{Comparison with Simulations} \label{sec:models_simulations}

\begin{figure*}
\epsscale{1.1}
\plotone{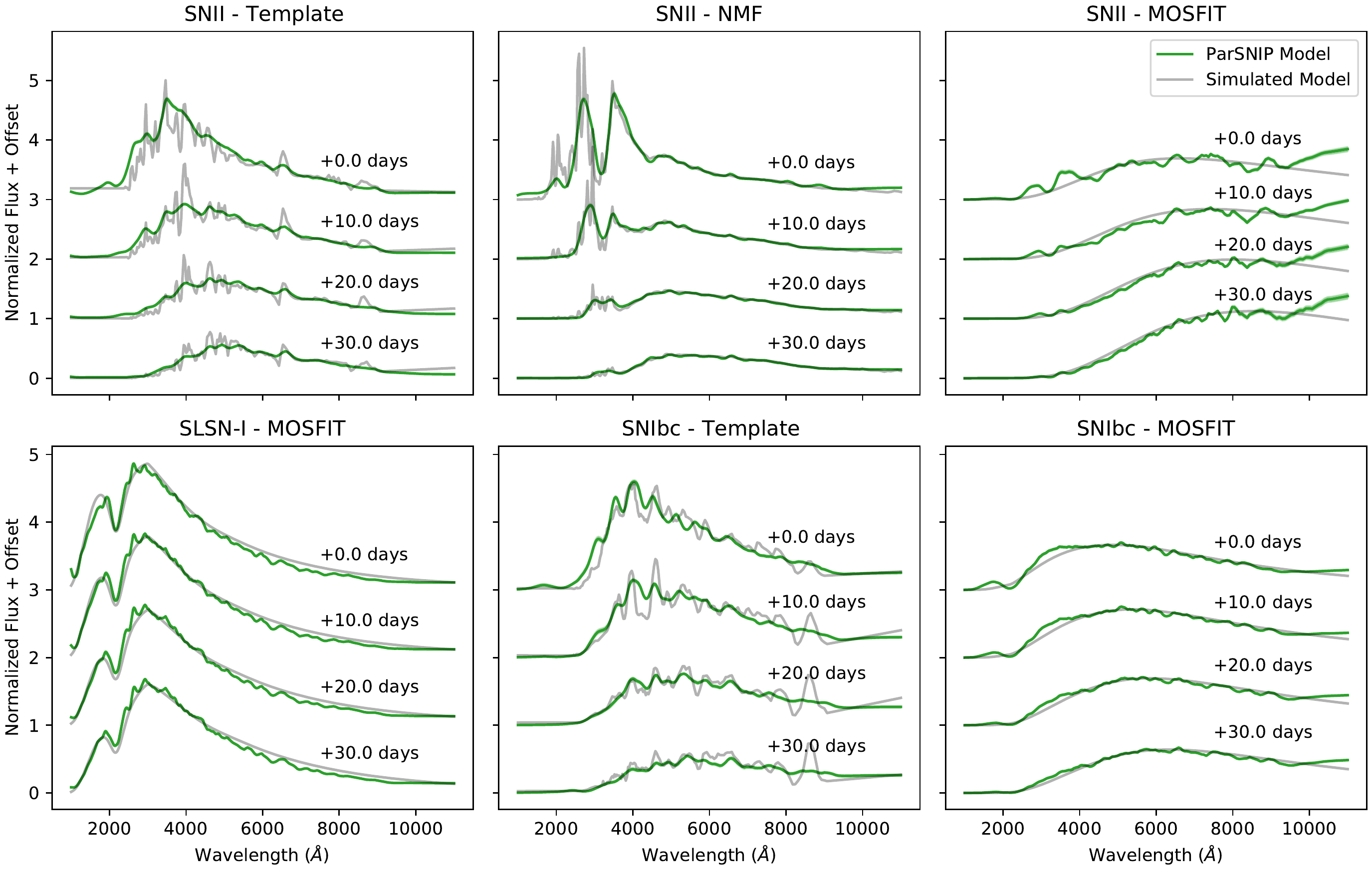} \\
\caption{
    Comparison of the spectra predicted by the ParSNIP model to the true spectra in the
    PLAsTiCC simulations. Each panel shows the simulated spectra for a given transient in
    grey at a range of different times. The spectra predicted by the ParSNIP model for that
    transient are shown in green. The titles of each panel
    indicate the type of transient and model that was used for the simulation.
    We find that the ParSNIP model is able to accurately recover the spectral time series
    of all of these different kinds of transients despite only being trained on photometry.
}
\label{fig:plasticc_spectra}
\end{figure*}

The PLAsTiCC dataset was simulated, so we can compare the spectra predicted
by the ParSNIP model directly to those that were used in the simulation. The result
of this procedure is shown for well-measured light curves of a range of different
types of transients in Figure~\ref{fig:plasticc_spectra}. Some classes, such as Type~II
supernovae, were simulated using several different models with very different spectral
properties. We find that the ParSNIP model is able to identify the different models
and recover the underlying spectral time series for each of them.

The spectra that are recovered by the ParSNIP model tend to be overly smooth compared
to the input simulations. This is primarily due to the fact that
we are learning the spectra by deconvolving photometry. The regularization term described
in Section~\ref{sec:spectra_regularization} can be adjusted to recover more of the
spectral features, but this comes at the cost of introducing additional noise into the model.
In practice, we find that they choice of regularization term has little impact on the
final photometry because all of the high frequency information is lost after the convolution
with the bandpass.

\section{Applications} \label{sec:applications}

There are many different applications for a generative model of all transient light curves.
We discuss three of them in this section. In Section~\ref{sec:classification}, we show
how the ParSNIP model can be used to perform photometric classification even with
heavily biased training sets. In Section~\ref{sec:novel}, we demonstrate a novel
method of searching for new kinds of transients. In Section~\ref{sec:distances},
we show how the ParSNIP model can be used to estimate cosmological distances to SNe~Ia.

\subsection{Photometric Classification} \label{sec:classification}

Upcoming surveys such as the LSST with the Rubin Observatory will
obtain photometric observations of millions of transients, but will only have the
spectroscopic resources to obtain spectroscopy of and label a small fraction of these transients.
Traditional classification techniques only use this small dataset of labeled transients for training
(e.g. \citet{lochner16, boone19}). In contrast, autoencoder-based methods can learn a low-dimensional
representation from the much larger dataset of both labeled and unlabeled transients.
This representation is a set of very informative features which can be used to
train a photometric classifier on the small labeled dataset.
This approach has previously been demonstrated in \citet{pasquet19} and \citet{villar20}.

The main advantage of ParSNIP over previous autoencoders is that the
representation that it learns was constructed to disentangle intrinsic
properties of transients from properties describing how transients
were observed. In particular, the intrinsic representation is disentangled
from redshift. The labeled training sets tend to be heavily
biased towards low-redshift transients, so redshift is not a good feature to use
for classification. \citet{pasquet19} attempted to construct a similar
representation using a contrastive loss function that adds a penalty
when transients with the same label have very different representations. This
technique is effective, but requires a representative sample of labeled 
high redshift transients that is often not available. In contrast, ParSNIP can generate such
a representation with only the unlabeled dataset because it is a full
generative model.

To perform photometric classification with ParSNIP, we first augment each light
curve in the training set 100 times following the procedure in
Section~\ref{sec:augmentation}. Note that we do not perform redshift augmentation
as in \citet{boone19}, we only use augmentation to obtain light curves at a wide
range of signal-to-noises and observing conditions. We then perform inference
using the ParSNIP model to estimate the latent representations for all of the transients in the
augmented training set. We convert the amplitude measured by the ParSNIP model
to a pseudo-luminosity $L$ using the cosmological parameters from \citet{planck20}
with the following formula:
\begin{align}
    L = -2.5 \log_{10}(A) - \mu_{\text{Planck20}}(z) + 25
\end{align}

The offset of 25 comes from the fact that the input fluxes for both the ParSNIP
and PLAsTiCC datsets were measured on the AB system with a zeropoint of 25. We
refer to $L$ as a pseudo-luminosity because it contains an arbitrary offset
from the unknown zeropoint of the ParSNIP model. This offset will be identical for
all light curves with the same intrinsic representation.
We use the full ParSNIP representation as the features for our classifier, consisting
of the intrinsic latent variables $s_1$, $s_2$, $s_3$, the pseudo-luminosity
$L$, and the color $c$. We also include the predicted uncertainties on all of these
measurements and the uncertainty on the reference time for a total of eleven features.
We do not include redshift as an explicit feature as the redshift distributions
of the labeled and unlabeled datasets tend to be very different, although the redshift
is used to calculate the luminosity. 

We train a gradient boosted decision tree on these features using the \texttt{lightgbm}
package \citep{ke17}. When training this classifier, we weight each transient in the
training set so that the sum of weights for each label is equal and so that the average
weight across all transients is one. We use the default hyperparameter values from
\texttt{lightgbm} with a \texttt{multi\_logloss} objective function, except that we
set the \texttt{min\_child\_weight} parameter to 1000
to avoid overfitting the augmented light curves. We train the classifiers with 10-fold
cross validation: we train each classifier on 90\% of the data and evaluate the model on
the remaining 10\%. We repeat this procedure ten times to obtain ten separate classifiers
and out-of-sample predictions for all of the transients in the training set. To avoid data
leakage, we ensure that all augmentations of the same light curve are in the same fold
for this procedure. For the unlabeled dataset, we generate predictions by averaging over
the outputs of all of the classifiers.

\subsection{Photometric Classification on the PS1 Dataset}

\begin{figure}
\plotone{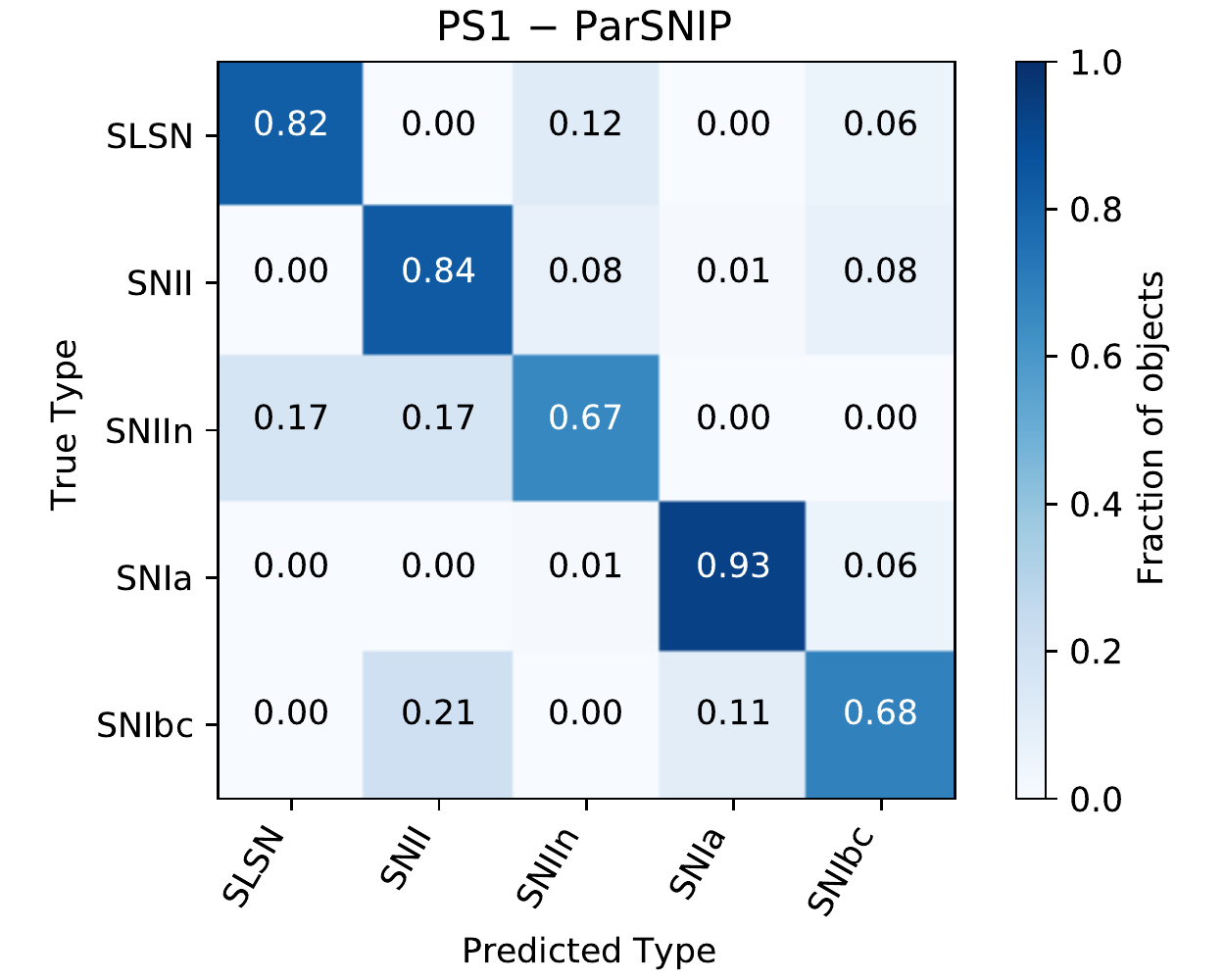}
\caption{
    Confusion matrix for the cross-validated ParSNIP predictions on the PS1 dataset.
    Each entry shows the fraction of transients of the type labeled on the left side of
    the plot that are assigned the corresponding type labeled on the bottom of the plot.
}
\label{fig:ps1_confusion}
\end{figure}

For the PS1 dataset, we only have labels for a subset of 557 transients. We predict
the labels for each of these transients using 10-fold cross-validation.
In Figure~\ref{fig:ps1_confusion}, we show a confusion matrix for this dataset.
Each row of the confusion matrix shows what fraction of the transients of a given
type are assigned to each label by the classifier. We find that the ParSNIP model is
able to accurately
classify all of the different types of transients in this dataset, with 63\% to 92\%
of the transients of each type being correctly identified in a five-way classification.

Our results are a significant improvement from both the SuperRAENN \citep{villar20} and
Superphot \citep{hosseinzadeh20} models that have previously
been applied to this dataset. We achieve an
overall accuracy of 89\% for the five-way classification compared to 87\% for SuperRAENN
on the same dataset. For the metric of the``macro-average completeness'' defined in
\citet{villar20} (the mean of the diagonal terms in the confusion
matrix), we achieve a value of 79\% with ParSNIP compared to 69\% for both SuperRAENN
and Superphot.  For a two-way classification of SNe~Ia compared to
all other labels, we
achieve a macro-averaged completeness of 96\% compared to 92\% for SuperRAENN.
This implies that a sample of SNe~Ia that was photometrically classified
with ParSNIP would have $\sim$2 times less contamination than a sample classified with
SuperRAENN.

These predictions were all made through cross-validation, so the classifiers were
evaluated on datasets that have very similar properties to the datasets that they
were trained on. The major advantage of ParSNIP over previous models is that its
representation is invariant to symmetries such as redshift, and it is designed
to perform well even when the training set differs significantly from the dataset
that the classifier is applied to. We expect that the improvement in performance
of ParSNIP over previous models such as SuperRAENN or Superphot will be even larger
when applied to the non-representative datasets such as the ones that will be produced
by the Rubin Observatory or the Rubin Space Telescope. Unfortunately it is impossible to test this
performance on real data since we do not have access to the true labels for the
higher redshift transients, although we can test it on simulations.

\subsection{Photometric Classification on the PLAsTiCC Dataset}

\begin{figure*}
\epsscale{1.15}
\plottwo{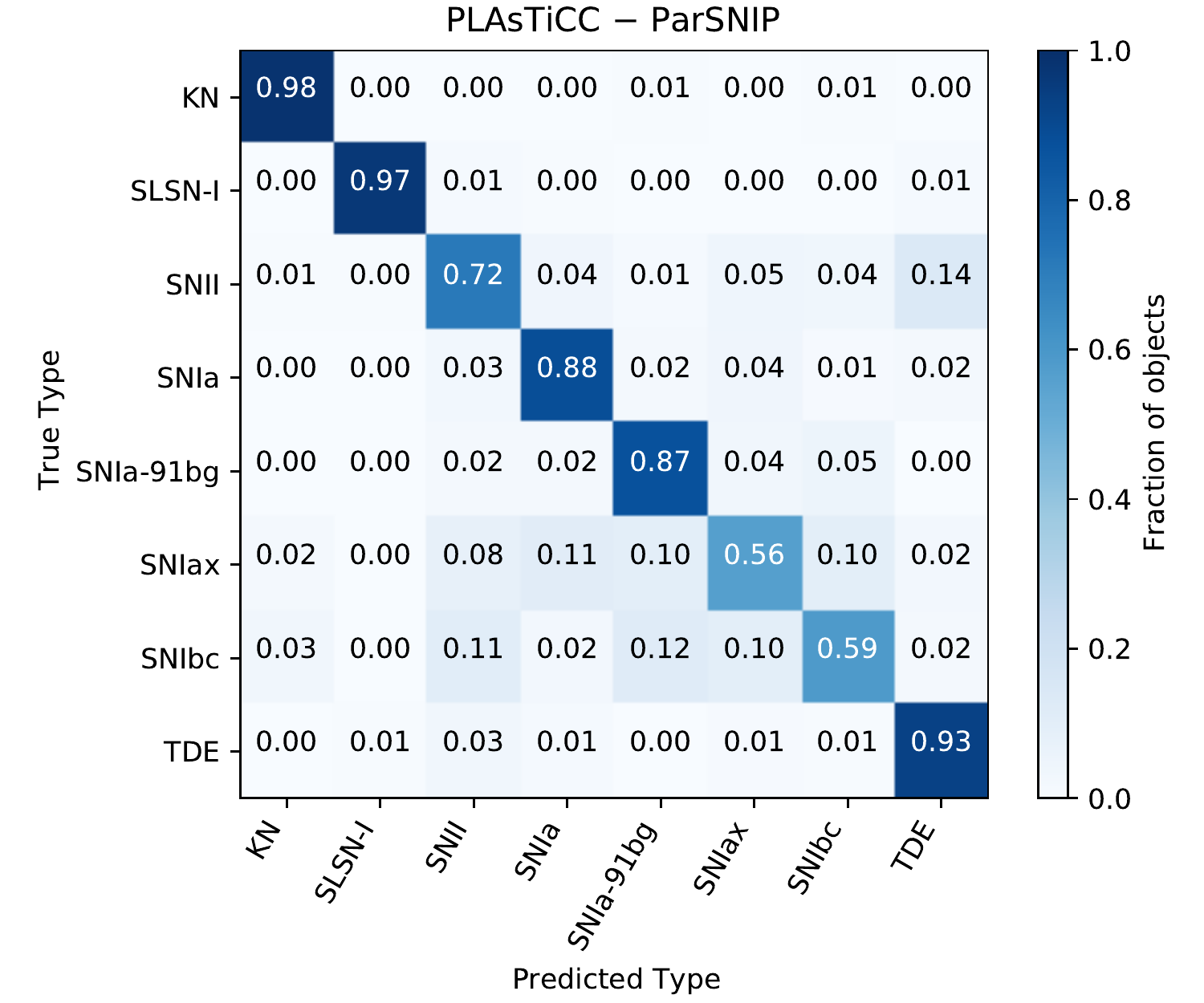}{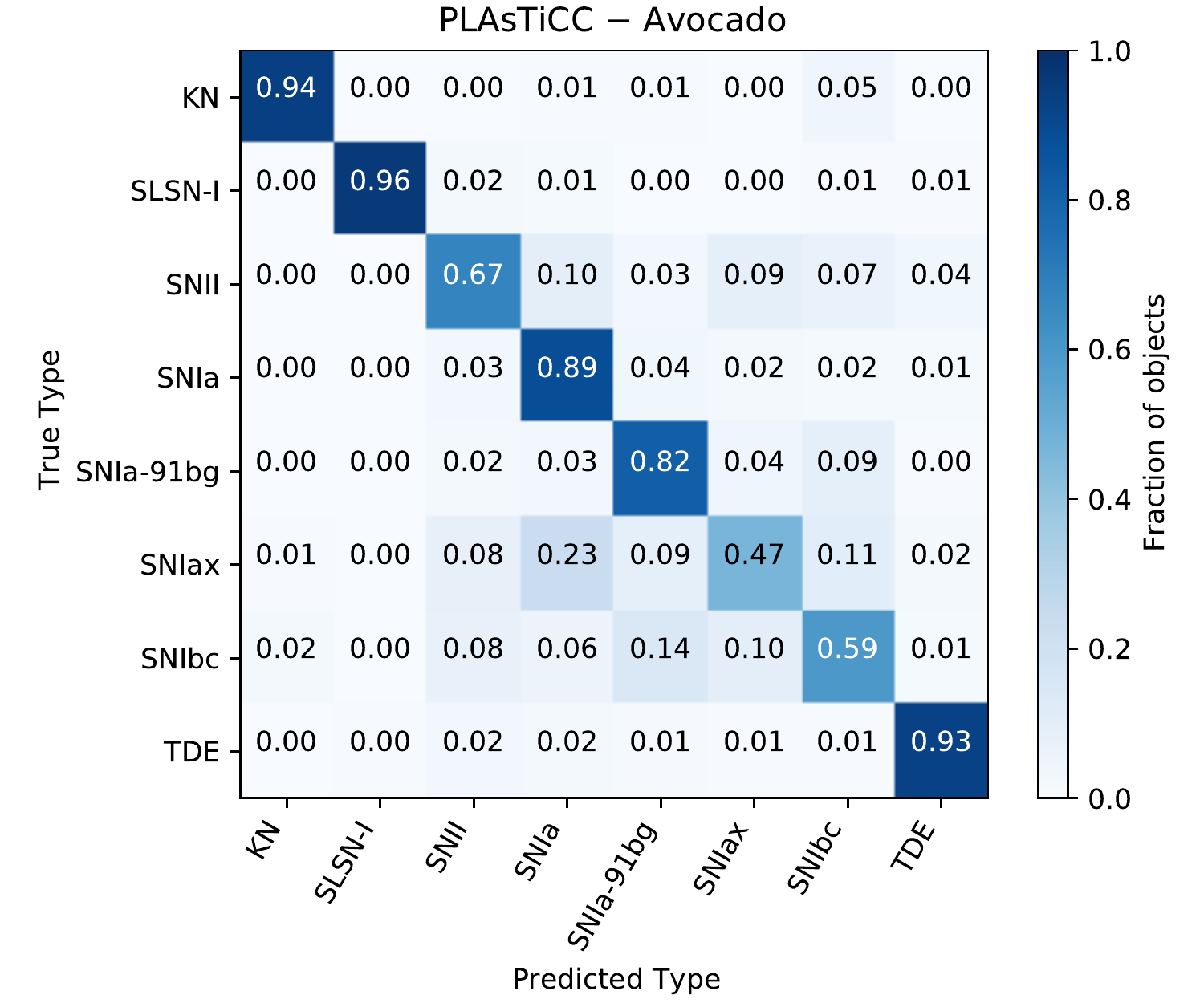}
\caption{
    Confusion matrices for the ParSNIP (left panel) and Avocado (right panel) predictions
    on the PLAsTiCC test set. Each entry shows the fraction of transients of the type labeled
    on the left side of the plot that are assigned the corresponding type labeled on the
    bottom of the plot. We find that the ParSNIP model has similar or better performance
    than the Avocado model for every kind of transient.
}
\label{fig:plasticc_confusion}
\end{figure*}

The PLAsTiCC dataset is a simulation of the LSST light curve sample, and has
a labeled training set that is highly nonrepresentative of the full dataset.
We compare the performance of the ParSNIP model for classification of the PLAsTiCC
light curves to the performance of the Avocado
model \citep{boone19} that was the best performing classifer in the PLAsTiCC challenge
\citep{hlozek20}. In the version of ParSNIP discussed in this work, we assume that
we know the redshift of each transient but this information was not available in
the original PLAsTiCC dataset. For a fair comparison, we retrained the Avocado
classifier using the true redshifts instead of photometric redshifts. We show the
confusion matrices for both the ParSNIP and Avocado classifiers in
Figure~\ref{fig:plasticc_confusion}.

\begin{figure}
\plotone{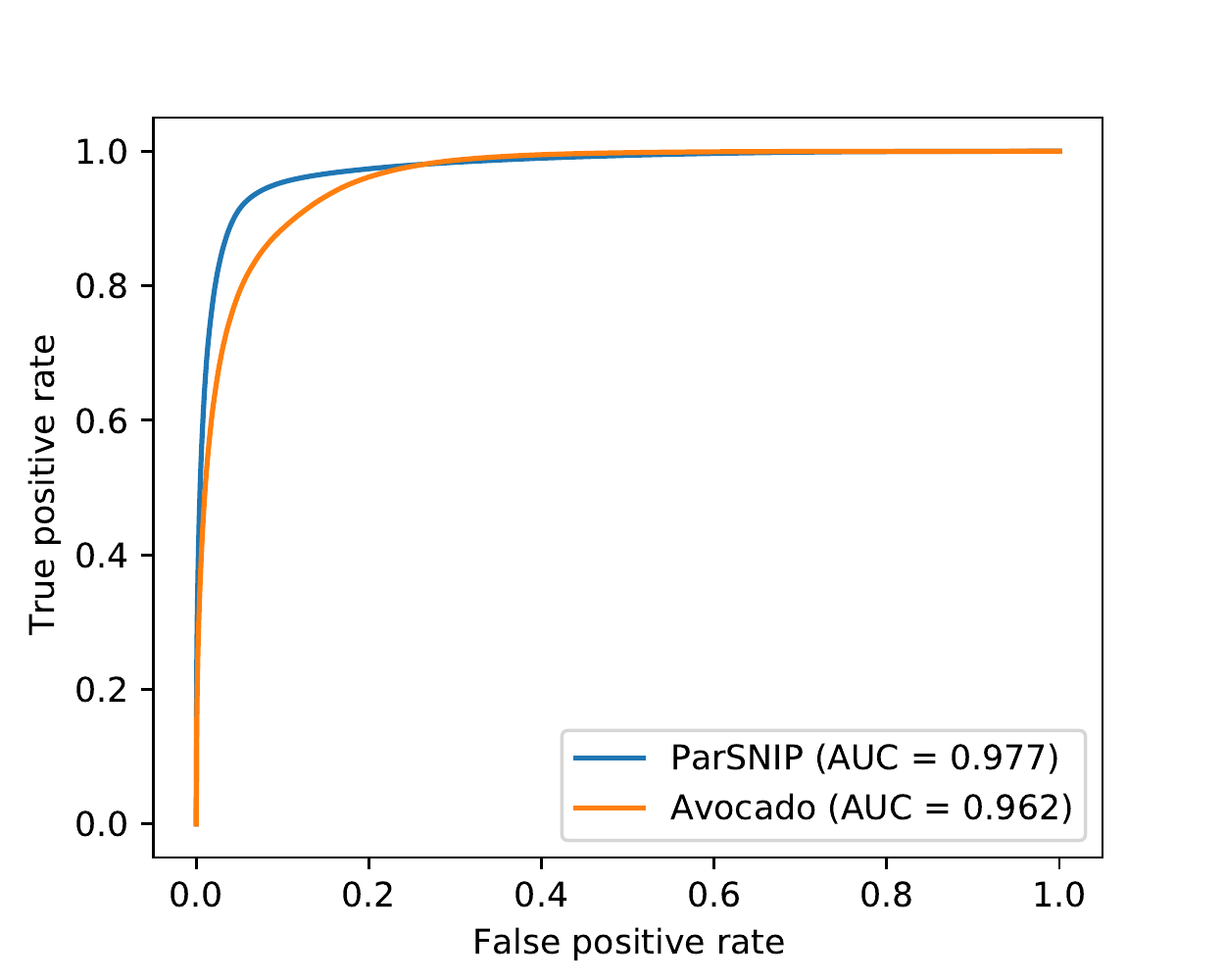}
\caption{
    Receiver operating characteristic (ROC) curve for SN~Ia classification
    on the PLAsTiCC dataset. The ParSNIP model (shown in blue) performs significantly
    better than the Avocado model (shown in orange), especially for thresholds corresponding
    to low false positive rates that are most interesting for supernova cosmology analyses.
}
\label{fig:plasticc_snia_roc}
\end{figure}

The ParSNIP model achieves similar or better performance than Avocado for
each of the different kinds of transients. We evaluated the weighted log-loss metric developed
for the PLAsTiCC challenge \citep{malz19} limited to the subset of transient types used in this
analysis. Avocado scores 0.599 on this metric while ParSNIP scores 0.535 which is a major
improvement. To illustrate how this difference in performance will affect astrophysical
analyses, we evaluated the receiver operating characteristic (ROC) curve for both of
these classifiers for SNe~Ia. This curve measures the
true positive rate (the fraction of SNe~Ia that are correctly
classified as SNe~Ia) as a function of the false positive rate (the fraction
of non-SNe~Ia that are misclassified as SNe~Ia) for different
thresholds of the classifier output. We show the ROC curve for both the ParSNIP and
Avocado classifiers in Figure~\ref{fig:plasticc_snia_roc}.

Measuring the Area Under the ROC Curve (AUC) (the integral of this curve), we find a value
of 0.977 for the ParSNIP model compared to 0.962 for the Avocado model. The improvement
of the ParSNIP model mainly comes from a major reduction in the false positive rate
for true positive rates below $\sim90\%$. For a true positive rate of 50\%, the ParSNIP
model has a false positive rate of only 0.44\% which is 2.3 times smaller than the false
positive rate of 1.04\% for the Avocado model. We find similar results for any true positive
rate below 0.9. This is of particular importance for cosmology analyses with SNe~Ia
where it is more important to have a clean dataset than include all transients
that were discovered. A SN~Ia cosmology analysis that uses the
ParSNIP model for classification will have 2.3 times less contamination from other
supernova types compared to an analysis that uses the Avocado model.

\begin{figure}
\plotone{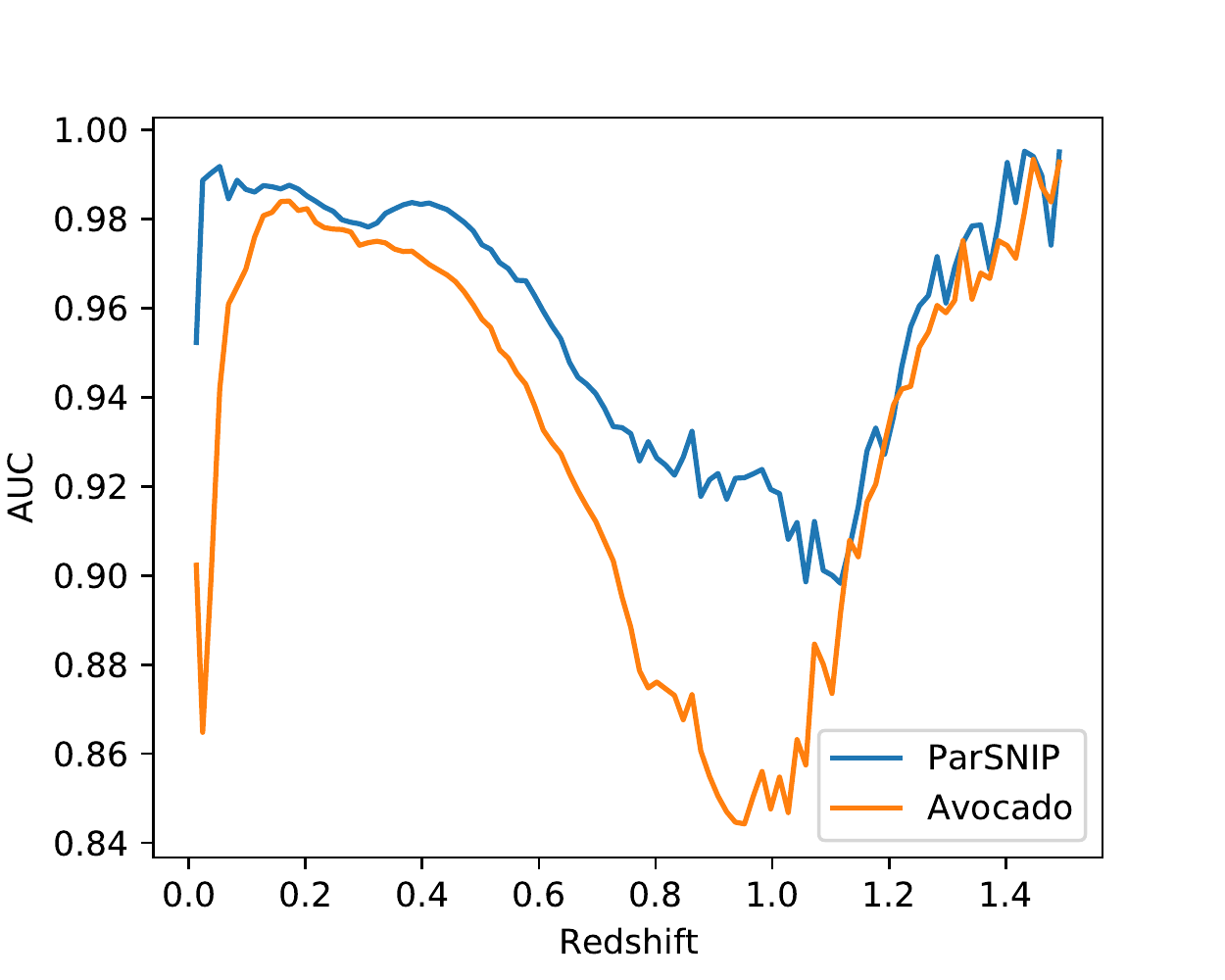}
\caption{
    AUC for classification of SNe~Ia as a function of redshift for both the
    ParSNIP model (shown in blue) and Avocado model (shown in orange) on the PLAsTiCC
    dataset. The ParSNIP model outperforms the Avocado model at all redshifts.
}
\label{fig:auc_redshift}
\end{figure}

Finally, we evaluated the performance of both the ParSNIP and Avocado models as a function
of redshift. We evaluated the AUC for classification of SNe~Ia in 100
evenly-spaced redshift bins between redshifts of 0 and 1.5. The results for both the Avocado
and ParSNIP models are shown in Figure~\ref{fig:auc_redshift}. We find that the ParSNIP
model is stable across different redshifts, and outperforms the Avocado model at all
redshifts. Of particular note is the decrease in performance for the Avocado model at very
low redshifts where we would expect to have very high signal-to-noise and well-measured
light curves. The Avocado model attempts to augment the training dataset by simulating
light curves at redshifts that are more representative of the full dataset, but struggles
to produce good simulations of very low redshift transients because light curves at those
redshifts are much higher signal-to-noise than most of the light curves in the training
set. Fundamentally, the Avocado augmentation process must be heavily tuned for each dataset
that it is applied to in order to minimize differences between the training and test sets.

In contrast, the ParSNIP model builds a redshift-independent representation of
light curves from the full dataset, and is mostly agnostic to differences in the
redshift distributions between the training and full datasets. Other than definitions
of the bandpasses, there are no instrument-specific elements of the ParSNIP model.
The same ParSNIP model was able to fit both the PLAsTiCC and PS1 datasets. We find
that the ParSNIP model has good performance at all redshifts without the need for
procedures such as redshift augmentation.

\subsection{Detecting Novel Transients} \label{sec:novel}

Upcoming surveys such as LSST will produce samples of transients that are more than
two orders of magnitude larger than
current samples, and could contain entirely new kinds of ``novel'' transients that have
never been previously observed. Previous work on detecting novel transients has typically
focused on identifying transients that are in some way different from the bulk of the
sample \citep{pruzhinskaya19, ishida19, martinezgalarza21, villar21}
without considering what was previously known about transients. As
a result, these methods typically produce a sample of transients that consists
mostly of rare transients (e.g. superluminous supernovae) rather than ones that have
truly never been observed.

In this work, we instead explore an alternative approach to anomaly detection enabled
by the ParSNIP model that takes prior knowledge into account. The ParSNIP model
was constructed so that transients with the same intrinsic properties
are embedded at the same location in the intrinsic latent space regardless of how
they were observed. Assuming that we have known labels for some subset of the transients in a
given dataset, we can identify novel transients by finding unlabeled transients
whose locations in the ParSNIP latent space are different from those of the transients
in the labeled subset. Note that this approach cannot be used with most
previously-used feature extraction methods because their features are correlated with
properties of the observations such as redshift.

We demonstrate this approach on the PLAsTiCC dataset.
The authors of the PLAsTiCC dataset included several different kinds of transients in the full
PLAsTiCC dataset that were not present in the labeled training set, including pair-instability supernovae
(PISN), intermediate luminosity optical transients (ILOT), and calcium-rich transients (CaRT)
\citep{kessler19}. As can be seen in Figure~\ref{fig:representations}, the PISNe and ILOTs
are well-separated from all other kinds of transients in the ParSNIP latent space.
We generate an augmented version of the PLAsTiCC training set with 100
realizations of each light curve following the procedure described in Section~\ref{sec:augmentation}
to cover a wide range of different observing conditions. For each transient in the full
dataset, we then calculate the Euclidean distance between the ParSNIP intrinsic latent variables
$\bs_i$ of that transient and the nearest transient in the augmented training dataset.
This distance measure can be interpreted as a novelty score, where transients with large
distances are very different from all of the transients in the training set.

Of the top 100 transients ranked by this novelty score, 87 are ILOTs and 3 are PISNe,
meaning that 90\% of the transients in 
this sample of ``novel transients'' are in fact a new kind of transient that isn't present
in the labeled training set. For comparison, \citet{villar21} use an isolation forest
on the same dataset. They obtain a sample of rare transients that is 95\% pure according
to their definition, but the vast majority of these rare transients are superluminous
supernovae, examples of which are available in the training set, and only $<5$\% of the transients
in their sample come from the types that are not included in the training dataset. We
stress that these are two entirely different approaches to anomaly detection. Which one is
more applicable will depend on the science goals of a given project. However, the ParSNIP representation enables
alternative approaches to anomaly detection that can take prior knowledge into account.

The methodology that we demonstrated here is very simple, and we only considered
the intrinsic ParSNIP representation. More advanced techniques could also
take advantage of luminosity, color, and redshift information or even additional information
such as properties of the host galaxies. They could also look at the density of
transients in each of the training and full datasets.
In practice, the detection of novel transients should involve a feedback loop where
resources are dedicated to followup candidates, and where these are then added back
into the training set. This form of ``active learning'' is discussed in detail in
\citet{ishida19} and an application to isolation forests for anomaly detection is
discussed in \citet{ishida21}. The ParSNIP model is ideally suited to being
used with these techniques.
It extracts a robust representation of transients in an unsupervised
manner, so it can be trained on unlabeled datasets containing novel transients.
Furthermore, the representation only consists of a small number of parameters,
so it is very computationally efficient to use the ParSNIP representation instead
of other feature extraction methods that can produce tens or hundreds of features.

\subsection{Estimating Distances to SNe~Ia} \label{sec:distances}

One major application of the SALT2 model is estimating distances to transients.
As a similar generative model, the ParSNIP model can also be used for distance estimation.
SALT2 and ParSNIP both estimate the brightnesses of transients relative to some arbitrary
zeropoint that can vary over the latent space. By subtracting off the zeropoint, we
obtain a distance estimate. For SALT2, the zeropoint is typically modelled using a linear
function of the SALT2 $x_1$ and $c$ parameters resulting in the following model for
estimating distances
\begin{equation}
    \mu_{\textrm{SALT2}} = M + \alpha x_1 + \beta c
\end{equation}
For ParSNIP, we use a similar model with linear corrections for all of the intrinsic
latent variables
\begin{equation}
    \mu_{\textrm{ParSNIP}} = M + \alpha_1 s_1 + \alpha_2 s_2 + \alpha_3 s_3 + \beta c
\end{equation}

We fit for the values of $M$, $\alpha_i$ and $\beta$ in both of these models by
comparing the estimated distance moduli to the distance moduli from the
corresponding redshifts using the cosmological parameters from \citet{planck20}.
In both cases, we use the sample of PS1 SNe~Ia with good SALT2 fits described
in Section~\ref{sec:salt_comparison}. We use Chauvenet's criterion to reject
SNe~Ia that are large outliers for either model which removes 8 of the 265
SNe~Ia in our sample. For the same set of SNe~Ia, we find that the RMS of the
SALT2 distance estimates is $0.155 \pm 0.008$~mag and the RMS of the ParSNIP
distance estimates is $0.150 \pm 0.007$~mag. When looking at
only the core of the distribution, we find that ParSNIP is able to estimate distances
with an NMAD of $0.126 \pm 0.009$~mag compared to $0.139 \pm 0.011$~mag for SALT2.
Hence, the ParSNIP model can
be used to estimate accurate distances to SNe~Ia, and it performs slightly better than
the SALT2 model for distance estimation on this sample of SNe~Ia.

One important caveat is that precision cosmology analyses
require a thorough understanding of light curve modeling uncertainties to avoid biases.
SALT2 includes an explicit description of the model covariance at every time/wavelength
that can be used to identify regions of the spectrum where the model is unreliable, such
as the UV/IR as seen in Figure~\ref{fig:snia_spectrum_comparison}. The ParSNIP model
uncertainties for Type~Ia supernova light curves should be thoroughly investigated and
understood before using ParSNIP in precision cosmology analyses.

\section{Discussion} \label{sec:discussion}

\subsection{Computational Requirements} \label{sec:computational_requirements}

All of the training and inference in this work was done using an NVIDIA GeForce
RTX 2080 Ti GPU. Training a new model takes approximately 2 hours for the PS1 dataset
and 23 hours for the PLAsTiCC dataset. The training time could also be decreased
significantly by adjusting the learning rate scheduler without having a major impact
on model performance. Inference takes approximately 11 seconds for
the PS1 dataset or 2 hours for the full PLAsTiCC dataset of over 3 million
light curves. The model can also be trained and evaluated on a CPU. When running
on 8 cores of an AMD Ryzen Threadripper 3970X, we find that the performance is
roughly 6 times slower than the numbers quoted previously for the GPU.

These computational times are not prohibitive even for large surveys
such as LSST. With a single GPU, the ParSNIP model could be retrained on a near
nightly basis and used to perform inference on the entire dataset of LSST light curves.
We can perform inference at a rate of approximately 500 LSST light curves per
second with a single GPU, so inference could easily be performed live on the
LSST alert stream.

\subsection{Supernova Cosmology Without Classification} \label{sec:cosmology_classification}

As discussed in Section~\ref{sec:classification}, surveys such as LSST will
rely on photometric classification to identify samples of SNe~Ia. Previous
cosmological analyses with photometrically-classified
SNe~Ia have used one model to identify whether a transient is an SN~Ia, and a second
model to estimate the distance to that transient assuming that it is an SN~Ia
\citep{hlozek12, jones18}. As a result, the distance estimates for any non SNe~Ia
that leak into the sample are heavily biased.

The ParSNIP model can be used both for classifying transients (as discussed in
Section~\ref{sec:classification} and estimating distances (as discussed in
Section~\ref{sec:distances}), so these two steps could be done simultaneously
in a supernova cosmology analysis. As part of the cosmology fit, one would
fit for the distance modulus zeropoint across the ParSNIP latent
representation as we did for the SN~Ia subset in Section~\ref{sec:distances}.
Fitting distances to all transients will require a more complex zeropoint model
than the simple linear one that we used for SNe~Ia because some transients have much
larger ranges of dispersion in luminosity.

Given such a zeropoint model, distances estimates to individual transients could be obtained
by marginalizing over the latent representation using the posterior distribution
from the ParSNIP model. This procedure would effectively
marginalize over all of the different kinds of transients that each light curve
is compatible with, and would estimate distances to all transients, not just
SNe~Ia. This procedure would reduce many of the biases present in current
cosmology analyses with photometrically-classified supernovae because
it estimates distances to non-SNe~Ia using a model trained on those
transients rather than a model trained on SNe~Ia. We plan on developing
such a model in future work.

\subsection{Photometric Redshifts} \label{sec:photoz}

The current implementation of the ParSNIP model relies on spectroscopic redshifts.
These redshifts will not be available
for all transients, and this is especially true when surveys are in progress.
In principle, the redshift could be treated as an additional explicitly-modeled
latent variable in the VAE. Upcoming
experiments such as the Rubin Observatory or the Roman Space Telescope
will have photometric
redshift models for the host galaxies of most of the transients they observe,
and these could be used as a prior for the VAE. This kind of analysis
would rely on understanding the posteriors of the photometric redshift models
and would have to be tuned to specific experiments.

We also find that the encoder is not very sensitive to changes
in the input redshift. We performed inference on the PS1 dataset with
noise added to all of the input redshifts with a standard deviation of
0.05~mag. The recovered intrinsic representation coordinates
change with an RMS of 0.20, 0.10, and 0.11 for $s_1$, $s_2$, and $s_3$
respectively. These differences are negligible for applications such
as photometric classification. Photometric redshifts could therefore
likely be used directly as inputs to a pretrained ParSNIP model
without a significant degradation in performance.

\subsection{Latent Variable Priors} \label{sec:priors}

The ParSNIP model learns priors on the intrinsic latent variables, but we
assumed fixed weak priors on the explicitly-modeled latent variables. We made this
assumption because the light curve datasets that the model was trained on
are highly biased. Astronomical surveys have some limiting magnitude
beyond which they will not detect transients, so the populations of
observed transients are biased towards brighter, bluer, and lower-redshift
transients. By using a fixed weak prior, we make no assumptions about these
biases and they can be corrected for in further analyses if necessary.

The explicitly-modeled latent variables tend to be very well
constrained. For the PS1 dataset, the median posterior uncertainty
on the reference time is 0.7 days compared to the prior of 20 days,
the median uncertainty on the color is 0.049 compared to the prior
of 0.3, and the median uncertainty on the amplitude is 3\%. All
of these variables are well constrained even for low signal-to-noise
light curves, so the choice of prior has minimal impact on the
posterior distributions. Applications such as photometric classification
or distance estimation only depend on the posterior distributions, so
they will not be significantly impacted by different choices of the
prior distribution.

Simulating new light curves from the ParSNIP model is one application
that does require knowledge of the prior distributions of the explicitly-modeled
latent variables. These prior distributions could be estimated
directly from a large dataset of light curves by explicitly
modeling the selection functions, although this would be very challenging.
Alternatively, the prior distributions can simply be estimated from theoretical
or empirical models as was done for the PLAsTiCC simulations.

\subsection{Latent Variable Posteriors} \label{sec:posteriors}

The decoder of the ParSNIP model is a generative model with a well-defined
likelihood. The encoder model uses variational inference to approximate the posterior
distribution over the latent variables for a given light curve, but it is also possible
to evaluate this posterior distribution directly. In particular, optimizers can be used
to find the maximum-a-posteriori values of the latent variables for a given light curve,
and techniques such as Markov chain Monte Carlo (MCMC) can be used to sample from the
posterior distribution directly. These approaches are currently used in supernova
cosmology analyses with models such as SALT2, and provide much more detailed information
about the posteriors but tend to be very computationally intensive. We added an interface
for the ParSNIP model to the \texttt{SNCosmo} package \citep{barbary16c} that implements
several methods of fitting light curves and evaluating posteriors.

\begin{figure*}
\plotone{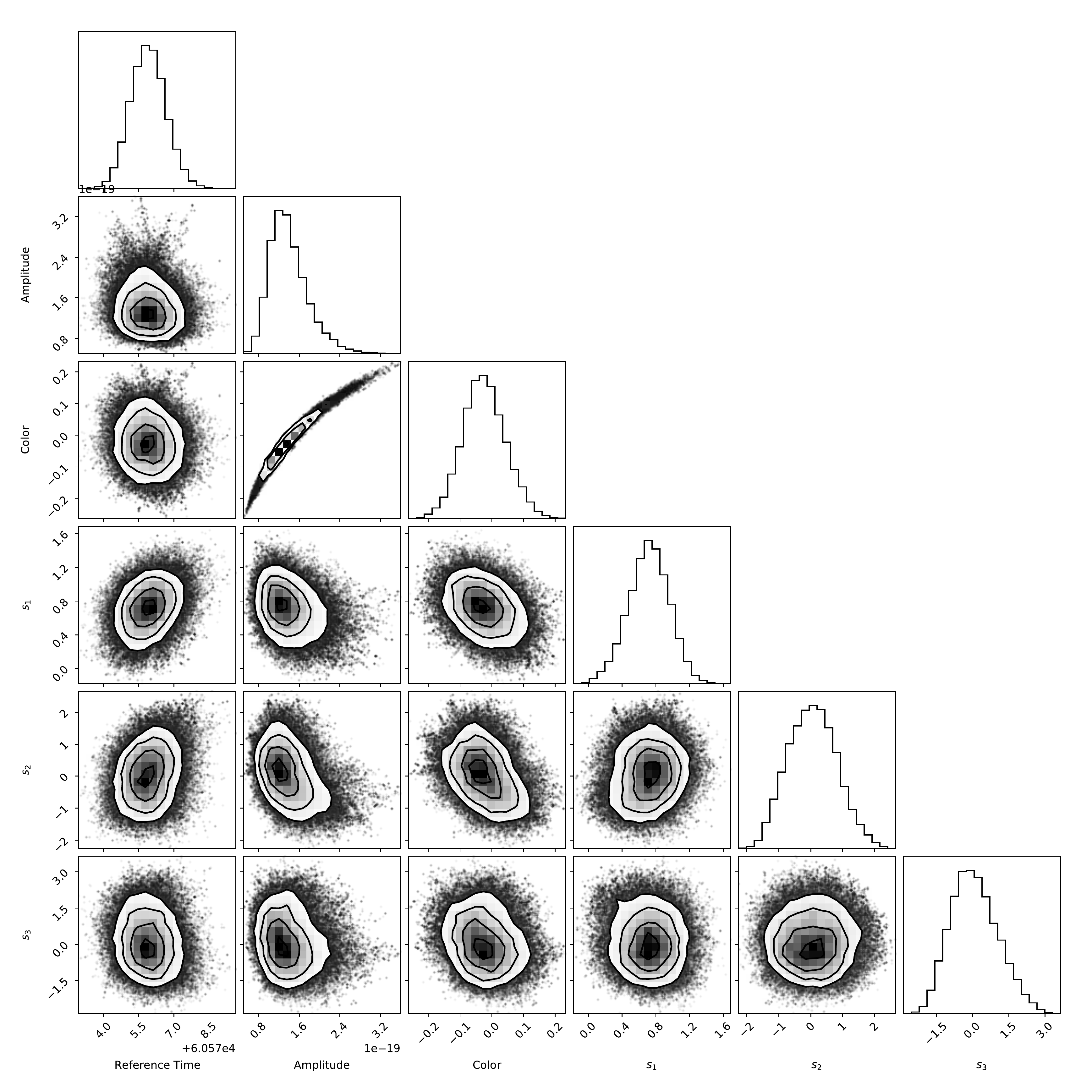}
\caption{
    Example of the posterior distributions of the latent variables in the ParSNIP model for a
    typical light curve in the PLAsTiCC dataset, sampled with MCMC. Each panel shows the distribution
    for two different latent variables, and the diagonal panels show the histograms of individual
    latent variables.
}
\label{fig:mcmc_posterior}
\end{figure*}

When training ParSNIP model using variational inference, we assumed that the posterior
of the latent variables can be described by a Gaussian distribution with a diagonal
covariance matrix. To test this assumption, we used the \texttt{emcee} \citep{foremanmackey13}
implementation of MCMC in \texttt{SNCosmo} to draw a large number of samples from the posterior
distributions of individual light curves. An example of the sampled posterior distribution for
a typical light curve from the PLAsTiCC dataset is shown in Figure~\ref{fig:mcmc_posterior}.

We find that the posterior distributions over the latent variables other than
amplitude are typically well-described by a Gaussian distribution with a diagonal
covariance matrix, and the
predicted means and standard deviations that we obtain from variational inference are
generally consistent with the true posterior distributions. This statement holds even for
light curves that are only partially sampled, and we do not see any evidence of multi-modal
posterior distributions. The amplitude and color latent variables are always highly correlated
which justifies our decision to marginalize over the amplitude as described in
Section~\ref{sec:marginalizingamplitude}. In general, we find that the posteriors estimated
by variational inference are reasonably accurate, but it may be desirable to evaluate the full
posteriors for applications such as supernova cosmology that require very high precision.

\subsection{Applying the ParSNIP Model to Other Kinds of Light Curves} \label{sec:other_lightcurves}

The model described in this paper was designed to be applied to short-lived
transient light curves, and should not be applied to other kinds of light curves without
modifications. In particular, the preprocessing procedure described in
Section~\ref{sec:preprocessing} roughly aligns and scales all of the light curves assuming
that they have a well-defined maximum. This is not a valid assumption for sources such as
variable stars or active galactic nuclei. We also subtract the background level
of each light curve using observations away from maximum light, and we constrain the model
to only output positive flux values. None of these assumptions are fundamental to the ParSNIP
model, and they should all be adjusted as appropriate if applying the model to other
kinds of light curves (e.g. not subtracting the background for variable stars).

\subsection{Including Additional Observations in the Training} \label{sec:including_spectra}

In this work we assumed that we only have photometry available for all of
the light curves in our sample. Large spectroscopic followup campaigns are
planned for most upcoming surveys. As an example, the 4MOST consortium
projects that they will obtain 30,000 spectra of live transients \citep{swann19}.
These spectra could be used to train the ParSNIP model. The ParSNIP model
already predicts the spectra at every phase when the model is compared to photometry,
and these model spectra could instead be compared to observed spectra directly.

When trained only on photometry, the ParSNIP model is required to learn
spectra by deconvolving the photometry. Including spectra of even a subset
of the transients would constrain the underlying spectral model. This
would be especially helpful for rare transients, such as PS1-12baa
discussed in Section~\ref{sec:models_other}, where there is limited
photometry of similar transients to constrain the positions and
widths of spectral features.

Additional followup photometry in different bands could similarly be included
to constrain the model at wavelengths where spectral coverage is typically lacking,
especially the UV and IR bands. Including a modest amount of IR photometry in the ParSNIP
training dataset would allow ParSNIP to learn a coarse IR spectral model. Such
models only currently exist for Type~Ia supernovae, but are essential for simulating
and planning upcoming experiments such as the Roman Space Telescope.

\subsection{Combining Light Curves from Multiple Surveys} \label{sec:combining_surveys}

Transient light curves have been collected by many different surveys, and the
ParSNIP model could be improved by training on a larger sample of light curves
observed by different instruments. The neural network in the decoder of the ParSNIP
model predicts spectra, and we explicitly compute photometry from those
spectra using the known bandpasses for an instrument. As a result, the ParSNIP
decoder can be used to predict the photometry for any instrument or bandpass
without any modification to the architecture or neural network weights.

The ParSNIP encoder, on the other hand, treats all of the individual bandpasses
separately. Nevertheless, it can operate on multiple instruments
by adding additional channels to the input representing the individual
bandpasses from each instrument. For instrument/bandpass combinations in which
a transient was not observed, we input zero at all times. With this approach,
the model can be trained simultaneously on light curves from different
surveys to produce a single latent representation and decoder model that
can be applied to all surveys. We tested this approach by training the
ParSNIP model simultaneously on the PS1 and PLAsTiCC datasets. We found
that the resulting model had a similar performance to the models trained
on the individual datasets, and transients with the same label in both
datasets are encoded at similar locations in the intrinsic representation.

\section{Conclusions}

The ParSNIP model developed in this work is a novel generative
model that describes how the spectra of transients evolve over time. This
model combines both a neural network to describe the intrinsic
diversity of transients and an explicit physics model of how light from
transients propagates through the universe and is observed.
It can be trained using only a large dataset of light curves without
requiring spectra or labels.

We demonstrated the effectiveness of our model both on both real (PS1) and simulated (PLAsTiCC) data.
With a three-parameter intrinsic representation, we are able to
reproduce out-of-sample light curves with model uncertainties of
$\sim$0.06~mag for the PS1 dataset and $\sim$0.04~mag for the PLAsTiCC
dataset. We also find that we are able to accurately predict the
spectra of the transients despite only training on photometry.
The ParSNIP model can estimate the distances to well-observed light curves
of SNe~Ia with an uncertainty of $0.150 \pm 0.007$~mag compared
to $0.155 \pm 0.008$~mag for the SALT2 model on the same sample of SNe~Ia.

Our model achieves state-of-the-art results
for photometric classification. It produces an intrinsic representation
that is independent of redshift, which means that it not highly sensitive
to the biases in training sets that are a major challenge for other methods.
For classification of SNe~Ia, the ParSNIP model produces
a sample that has 2 times less contamination compared to the SuperRAENN
model on the PS1 dataset, and 2.3 times less contamination compared to
the Avocado model on the PLAsTiCC dataset. The performance of the ParSNIP
model is stable across all redshifts and does not require the use
of techniques such as redshift augmentation. Our model
can also be used to detect novel transients. In a simulated dataset,
of 100 novel transients suggested by our algorithm, 90\% of them
are a new kind of transient that had never previously been observed.

All of the results in this paper can be reproduced with the publicly available
\texttt{parsnip} software package \footnote{\url{https://doi.org/10.5281/zenodo.5493509}}.
This package contains instructions for how to access all of the data that
was used in our analyses. It also contains Jupyter notebooks that can be
used to reproduce all of the figures in this paper.

\acknowledgments

We thank Andrew Connolly, Kara Ponder, Stephen Portillo, John Franklin Crenshaw,
Greg Aldering, Saul Perlmutter, and the anonymous referee for valuable
feedback and discussions.
K. B. acknowledges support from the DiRAC Institute in the Department
of Astronomy at the University of Washington. The DiRAC Institute is supported
through generous gifts from the Charles and Lisa Simonyi Fund for Arts and Sciences, and the Washington Research Foundation.

The Pan-STARRS1 Surveys (PS1) and the PS1 public science archive have been made possible through contributions by the Institute for Astronomy, the University of Hawaii, the Pan-STARRS Project Office, the Max-Planck Society and its participating institutes, the Max Planck Institute for Astronomy, Heidelberg and the Max Planck Institute for Extraterrestrial Physics, Garching, The Johns Hopkins University, Durham University, the University of Edinburgh, the Queen's University Belfast, the Harvard-Smithsonian Center for Astrophysics, the Las Cumbres Observatory Global Telescope Network Incorporated, the National Central University of Taiwan, the Space Telescope Science Institute, the National Aeronautics and Space Administration under Grant No. NNX08AR22G issued through the Planetary Science Division of the NASA Science Mission Directorate, the National Science Foundation Grant No. AST-1238877, the University of Maryland, Eotvos Lorand University (ELTE), the Los Alamos National Laboratory, and the Gordon and Betty Moore Foundation.


\vspace{5mm}

\facilities{
    PS1
}


\software{
    Astropy \citep{astropy13, astropy18},
    corner \citep{foremanmackey16},
    emcee \citep{foremanmackey13},
    Extinction \citep{barbary16b}
    % George \citep{ambikasaran15},
    Jupyter \citep{kluyver16},
    LightGBM \citep{ke17},
    Matplotlib \citep{hunter07},
    NumPy \citep{vanderwalt11},
    % Pandas \citep{mckinney10},
    PyTorch \citep{pytorch},
    scikit-learn \citep{scikit-learn},
    SciPy \citep{scipy},
    SNCosmo \citep{barbary16c}
}



% \appendix

% \section{Appendix information}


\bibliography{references}{}
\bibliographystyle{aasjournal}



\end{document}